\pdfoutput=1

\documentclass[fleqn,10pt]{wlscirep}
\usepackage[utf8]{inputenc}
\usepackage[T1]{fontenc}
\usepackage{booktabs}
\usepackage{wrapfig}
\fancypagestyle{preprintnotice}{%
  \fancyhf{}%
  \fancyfoot[C]{\footnotesize\textit{Preprint}}%
}
\providecommand{\linelabel}[1]{}

\title{Uncertainty-Aware End-to-End AI Weather Forecasting: Disentangling Observation and Model Contributions}

\author[1,*]{Rodrigo Almeida}
\author[1]{Noelia Otero}
\author[1]{Jost Arndt}
\author[1]{Simon Baur}
\author[1,2,3]{Wojciech Samek}
\author[1]{Jackie Ma}

\affil[1]{Applied Machine Learning Group, Fraunhofer Heinrich-Hertz Institute, 10587 Berlin, Germany}
\affil[2]{Department of Electrical Engineering and Computer Science, Technische Universität Berlin, 10587 Berlin, Germany}
\affil[3]{BIFOLD - Berlin Institute for the Foundations of Learning and Data, 10587 Berlin, Germany}
\affil[*]{rodrigo.almeida@hhi.fraunhofer.de}


\begin{abstract}
    End-to-end weather forecasting systems produce skillful global gridded and station forecasts directly from raw Earth observations, replacing the numerical weather prediction pipeline, including data assimilation, at a fraction of its cost.
    These systems are deterministic and issue no uncertainty.
    Here we render the Aardvark Weather model probabilistic by attaching one stochastic mechanism to each component: learned, input-dependent noise at the observation encoder, aimed at the uncertainty inherited from the observing system, and Monte Carlo dropout in the processor, aimed at the uncertainty of the learned dynamics.
    The resulting nested ensemble attributes forecast spread to the two sources through a law-of-total-variance decomposition, cross-checked by withholding observation streams.
    Probabilistic finetuning significantly improves the mean forecast, by 11\% on average across variables and lead times.
    The ensemble is close to calibrated against ERA5 through the medium range, is at least as accurate as the deterministic model at weather stations while beating it in relative economic value at every lead time, and remains considerably behind the
operational ECMWF IFS ensemble.
    The encoder branch behaves consistently with observation-driven uncertainty.
    Component-attributed uncertainty makes end-to-end forecasts more transparent, a step toward observation-driven digital twins of the atmosphere.
\end{abstract}
\begin{document}

\flushbottom
\maketitle
%
%
\thispagestyle{preprintnotice}

%

\section*{Introduction}


Weather forecasting has improved over recent decades in a quiet revolution,
driven by a deepening understanding of atmospheric physics and by numerical
weather prediction (NWP) models that solve the governing equations of the
atmosphere\cite{bauer2015quiet}. Deterministic, data-driven neural networks have
recently matched or surpassed these models in skill at a fraction of their
computational
cost\cite{pathak2022fourcastnet,bi2023pangu,lam2023graphcast,chen2023fengwu}.
They are, however, trained on and initialized from reanalysis datasets such as
ERA5\cite{hersbach2020era5}: gridded reconstructions of the atmosphere produced
by data assimilation, the process of merging scattered, irregular observations
with a background forecast\cite{carrassi2018da}. Such models therefore depend on
the assimilation pipeline, and using them directly with cycling assimilation
procedures is possible but
non-trivial\cite{slivinski2025assimilating,huang2026cycling}. End-to-end (E2E)
models remove this dependence by forecasting directly from raw observations.
Aardvark Weather\cite{allen2025aardvark} integrates satellite, station, ship and
balloon measurements into forecasts at weather station locations through a
modular pipeline: an encoder that assimilates the observation streams into a
gridded analysis, a processor that advances that state in time, and a decoder
that maps the forecast to station locations. Similar systems, notably ECMWF's
GraphDOP (direct-to-observations prediction)\cite{alexe2024graphdop}, are
likewise trained exclusively on Earth system observations. We build on Aardvark
in this work because its modular encoder--processor--decoder design allows
stochasticity to be introduced and attributed per component, and because its
model weights and training data are openly
available\cite{allen2025aardvark,anna_vaughan_2025}.

For weather forecasts, probabilistic output is required for rational
decision-making: the optimal action depends on comparing an event probability
with a user's own cost--loss ratio, which is different across users, so no single
deterministic forecast can be optimal for all of
them\cite{murphy1977costloss,fundel2019dialogue}. Operational assessment of AI
weather models accordingly stresses probabilistic skill and forecast value over
headline deterministic scores\cite{benbouallegue2024rise}, and AI ensembles have
recently been shown to deliver greater economic value than the leading NWP
ensemble\cite{price2025gencast}. Uncertainty can be introduced into an AI model
in many ways\cite{abdar2021review}; for weather models trained on reanalysis
this has been done through input
perturbations\cite{almeidaPredictiveSkillArtificial2026,buelte2024uq},
stochastic noise injection\cite{langAIFSCRPSEnsembleForecasting2026} and
diffusion models\cite{price2025gencast}, mirroring the initial-condition and
model perturbations through which classical NWP ensembles, which remain the
state of the art in calibrated probabilistic
forecasting\cite{leutbecher2008ensemble}, generate their
spread\cite{molteni1996eps,buizza1995singular,toth1997breeding,isaksen2010eda,bonavita2012eda,palmer2009stochastic,leutbecher2017stochastic}.
End-to-end models have so far been left out of this development. 
\begin{figure}[!tb]
	\centering
	\includegraphics[width=\linewidth]{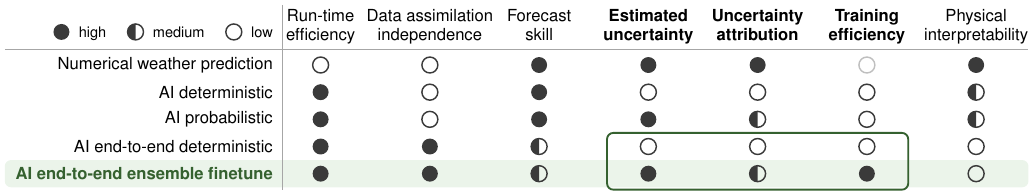}
	\caption{Qualitative comparison of weather forecasting systems. In bold the main contributions of this work are highlighted.}
	\label{fig:model-classes}
\end{figure}
Figure~\ref{fig:model-classes} summarizes the qualitative trade-offs between
these system classes: end-to-end models forecast directly from raw observations
at minimal cost, but issue neither an uncertainty estimate nor an
attribution of that uncertainty to its sources, the gap this work addresses.

Predictive uncertainty is commonly divided into two components: an aleatoric
part, inherited from the variability of the observed system and by definition
irreducible, and an epistemic part, originating in the model itself and in
principle reducible through additional data, training or better
parametrization\cite{derkiureghian2009aleatory,kendall2017uncertainties,huellermeier2021aleatoric}.
Estimating and disentangling the two is an active research field across machine
learning domains: in safety-critical applications such as medical imaging, the
split can inform whether an uncertain prediction calls for a better model or
must be deferred to a clinician\cite{baur2026disentanglement,baur2026beyond},
and it is equally consequential in AI weather and climate
modeling\cite{depeweg2018decomposition,valdenegro2022deeper,houlsby2011bald,smith2018understanding,mansfieldEpistemicAleatoricUncertainty},
with the resulting estimates verified using the proper scores and calibration
diagnostics standard in earth system
modeling\cite{haynesCreatingEvaluatingUncertainty2023,hersbach2000crps,gneiting2007scoring,gneiting2007calibration,hamill2001rank}.
Knowing whether forecast spread stems from the observations or from the model
tells developers and operational centers where improvement is possible.
Uncertainty inherited from the observing system points to the value of new or
better-placed observations, which can inform the design of future observation
networks, while epistemic uncertainty can be addressed with more training data,
better architectures or larger ensembles. For end-to-end models the question is
new because the model constructs its own atmospheric state from raw observations
rather than starting from a fixed analysis state.

\begin{figure}[!htb]
	\centering
	\includegraphics[width=\linewidth]{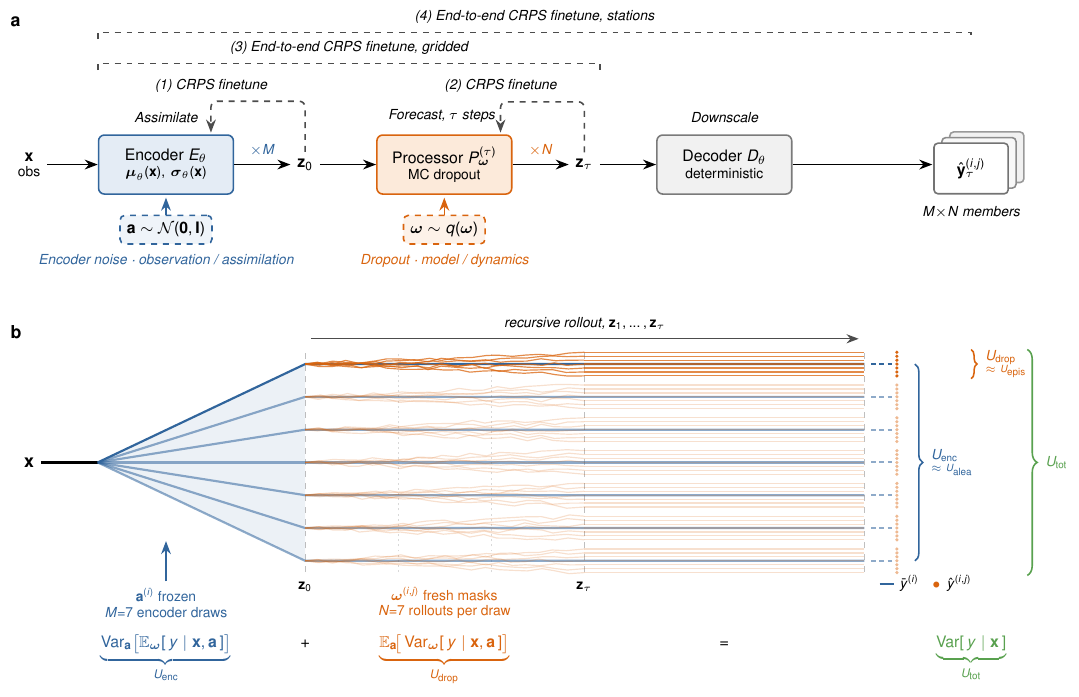}
	\caption{E2E ensemble architecture.
	\textbf{a}, The deterministic encode--process--decode pipeline uplifted to a
	stochastic map: learned, input-dependent Gaussian noise $\mathbf{a}$
	(blue)\cite{kendall2017uncertainties} enters at the encoder and Monte~Carlo
	dropout\cite{gal2016dropout} over the weights $\boldsymbol{\omega}$ (orange)
	perturbs the processor; the decoder remains deterministic, so
	each random source is attached to exactly one component. Both stochastic
	components are finetuned in turn against a fair-CRPS objective on their
	own output (dashed arrows): (1) the encoder's noise amplitude on the
	analysis $\mathbf{z}_0$, then (2) the processor, with dropout active,
	along the rollout (see \hyperref[sec:training]{Data and training}).
	The whole chain is then finetuned end-to-end against one of two targets (Supplementary Note~\ref{note:joint}): (3) for the gridded forecasts, encoder and
	processor are trained jointly;
	(4) for the station forecasts, encoder, processor and station
	decoder are trained jointly. Sampling
	the encoder noise $M$ times and the dropout masks $N$ times per draw
	yields an $M{\times}N$-member forecast ensemble.
	\textbf{b}, The nested ensemble, aligned with the pipeline
	above: each of the $M=7$ encoder draws $\mathbf{a}^{(i)}$ shifts the
	assimilated state $\mathbf{z}_0$, fanning the forecast into group spines
	(blue). These are the dropout-averaged conditional means $\bar y^{(i)}$. The $N=7$
	MC-dropout rollouts per encoder draw (orange; masks redrawn at every
	step of the recursive rollout; one group highlighted) scatter the members
	$\hat y^{(i,j)}$ around their spine, and the deterministic decoder (straight
	segments) adds no further spread. Within-group scatter estimates the
	dropout branch variance $U_{\mathrm{drop}}$ ($\approx$ model/dynamics) and
	between-group-mean scatter the encoder branch variance $U_{\mathrm{enc}}$
	($\approx$ observation/assimilation) of the law-of-total-variance split
	$U_{\mathrm{enc}}+U_{\mathrm{drop}}=U_{\mathrm{tot}}$, via unbiased ANOVA
	estimators (see \hyperref[sec:vardecomp]{Methods}).}
	\label{fig:method}
\end{figure}

The central methodological contribution of this work is a nested ensemble method
that uplifts a pretrained deterministic end-to-end model into a probabilistic
forecaster whose uncertainty is attributable to its components, without training
a new probabilistic system from scratch. We apply it to the pioneering Aardvark
Weather model\cite{allen2025aardvark} and refer to the uplifted system as the
E2E ensemble. Exploiting the modular architecture, we introduce stochasticity at
two points (Fig.~\ref{fig:method}): learned, input-dependent noise in the
encoder\cite{langAIFSCRPSEnsembleForecasting2026}, aimed at the uncertainty of
the observations and their assimilation, and Monte Carlo
dropout\cite{gal2016dropout} in the processor, aimed at the uncertainty of the
learned forecast dynamics. Both sources are finetuned on top of the
deterministic model separately and jointly, and then combined in a
nested ensemble, following a curriculum of deterministic pretraining and short
probabilistic finetuning recently shown to reach frontier probabilistic skill at
a fraction of the training cost\cite{cachay2026ucast}. Sampling the two sources
in a nested design, with an outer ensemble over encoder noise draws whose
members are each expanded by an inner ensemble over dropout masks, is what makes
the resulting spread decomposable into the two contributions.
\linelabel{rev:R1-M7}The aleatoric/epistemic reading of the two sources is a
design choice to be tested, so we name them after their model component
throughout, the encoder branch and the dropout branch, and return to their
interpretation once the attribution has been tested.

This design lets us pose three questions that reanalysis-trained probabilistic
systems cannot answer, because an end-to-end model constructs its own atmospheric
state: \textbf{(1) explanation:} \emph{where does the predictive uncertainty of
	an end-to-end model originate\cite{bley2025secondorder}: in the observation
	encoding (assimilation) step, the analogue of observation and analysis
	uncertainty in NWP, or in the learned forecast dynamics, the analogue of model
	uncertainty?} \textbf{(2) disentanglement:} \emph{can these two contributions be
	disentangled and separately attributed, in the aleatoric/epistemic sense of the
	deep learning uncertainty quantification
	literature\cite{kendall2017uncertainties,huellermeier2021aleatoric}?}
\textbf{(3) cost:} \emph{can stochasticity be finetuned cheaply into a
	pretrained end-to-end system without degrading the mean forecast?} Here we show,
verifying against ERA5\cite{hersbach2020era5}, the operational IFS ensemble and
HadISD station observations\cite{dunn2012hadisd}, that the probabilistic
finetuning improves rather than degrades the mean forecast, that the resulting
ensemble is close to calibrated across lead times, and that a
law-of-total-variance decomposition, cross-checked with an observation-denial
experiment, \linelabel{rev3:C2-e}attributes the spread to the encoder and
dropout branches, which approximate its observation and model sources. Such
credibility is necessary for adoption: stakeholders who act on probabilistic
forecasts, such as civil protection agencies, grid operators and insurers, need
to trust not only the forecast but also its stated confidence. A forecasting
system that can explain where its uncertainty comes from is more transparent,
easier to audit\cite{dramsch2025explainability}, and a step toward
observation-driven digital twins of the atmosphere, in which the effect of
adding, removing or degrading an observation stream on forecast confidence can
be assessed directly.

\section*{Results}

\subsection*{Probabilistic finetuning significantly improves the mean forecast, by 11\% on average}

Uplifting the Aardvark model to a stochastic map significantly improves the
forecast skill: the E2E ensemble mean beats the original configuration at every lead
time and for all six headline variables: 2-m temperature (T2M), 850~hPa
temperature (T850), the 10-m zonal and meridional wind (U10, V10), 500~hPa
geopotential (Z500), and mean sea-level pressure (MSLP), with RMSE (Root Mean
Squared Error, lower is better) reductions of up to $18\,\%$
(Figure~\ref{fig:skill}a); averaged over all $60$ variable--lead combinations
the reduction is $11.0\,\%$ (95\% CI $[-11.8, -10.2]\,\%$). Among all variables,
the ensemble mean is $8.0\,\%$ better at day~1 (95\% CI $[-9.2, -6.9]\,\%$) and
$13.6\,\%$ better at day~10 ($[-14.7, -12.6]\,\%$), the gain growing with lead
time from its day-1 minimum. The improvement is statistically significant for
all 60 variable--lead combinations, and no variable is degraded at any lead
time. \linelabel{rev:R1-M1} This improvement in the forecast skill can be
attributed to the averaging of sharp ensemble members. \linelabel{rev:R1-M2}
Judged using the relative economic value (REV) for high and low events (90th and
10th percentiles), the advantage is clearer: at the cost--loss ratio where REV
peaks, the E2E ensemble's REV exceeds the deterministic forecast's by $0.15$ at
analysis time and by $0.17$--$0.31$ at forecast lead times in the variable mean,
on a scale where $0$ is climatology and $1$ a perfect forecast
(Figure~\ref{fig:skill}b). The margin widens with lead time because the
deterministic forecast's value fades faster: by day~10 its maximum REV has
fallen to $0.07$ while the ensemble retains $0.32$. The probabilistic output
comes on top of, not instead of, deterministic skill. Compared with the
operational IFS ensemble the end-to-end model remains behind in probabilistic
skill at all leads, with the CRPS gap largest at short leads ($71\,\%$ in the
variable mean at day~1) and narrowing to $9$--$17\,\%$ by day~10
(Figure~\ref{fig:skill}c). The gap is largest for the mass fields (Z500 and
MSLP). In operational systems these fields are constrained by radiosonde,
aircraft, and radio-occultation observations\cite{bormann2019ose}. The encoder
receives no aircraft or radio-occultation data at all, and its radiosonde stream
carries little weight (it is the least influential stream in the observing
system experiments below). This points to initial-condition error inherited from
the original Aardvark analysis\cite{allen2025aardvark}.

\subsection*{The E2E ensemble is close to calibrated across lead times}

\begin{figure}[!htb]
	\centering
	\includegraphics[width=\linewidth]{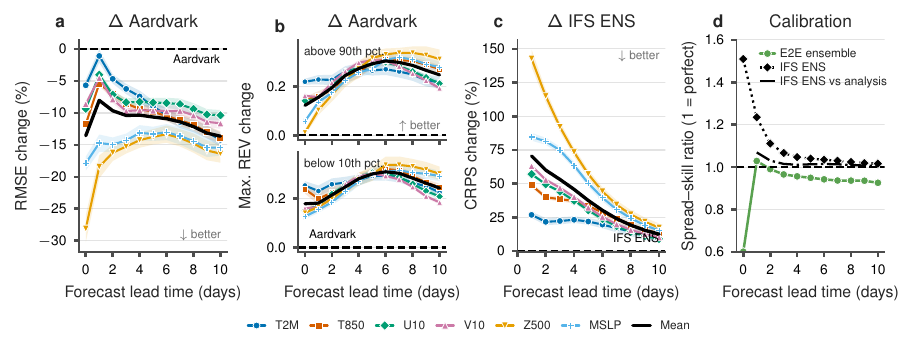}
	\caption{Skill and calibration of the E2E ensemble on the grid.
		\textbf{a}, Relative change in the ensemble-mean RMSE with respect to
		Aardvark per headline variable; negative is better, the black line is
		the mean over variables and the dashed line marks Aardvark.
		\textbf{b}, Difference in maximum relative economic value (REV)
		between the E2E ensemble and Aardvark, positive where the ensemble is
		more valuable, for events above the local 90th (top) and below the
		10th (bottom) climatological percentile.
		\textbf{c}, E2E ensemble CRPS relative to the operational IFS ENS;
		positive values mean the E2E ensemble is worse, and the zero line marks
		IFS ENS. \textbf{d}, Spread--skill ratio (SSR) of the E2E ensemble
		averaged over the headline variables (per-variable curves in
		Supplementary Fig.~\ref{fig:supp-ssr}), with IFS ENS verified against
		ERA5 (solid) and against its own analysis (dash-dotted) for reference;
		the dashed line marks perfect calibration.}
	\label{fig:skill}
\end{figure}

\linelabel{rev:R1-M3}The E2E ensemble is close to calibrated at every forecast
lead time: the variable-mean spread--skill ratio is $1.03$ at day~1 and drifts
slowly towards mild under-dispersion, $0.95$ at day~5 and $0.93$ at day~10,
whereas IFS ENS starts over-dispersive at day~1 ($1.24$) and settles just above
one (Figure~\ref{fig:skill}d). At analysis time (lead time 0) the ensemble is
the seven-member encoder output and all of its spread comes from the encoder
branch. There it is under-dispersive ($\mathrm{SSR}=0.60$ in the variable mean,
from $0.68$ for V10 down to $0.43$ for Z500), \linelabel{rev:R2-2} which points
at limitations in the noise injection in the encoder branch. Calibration is
improved within the first forecast step, once the dropout branch is present: 
the dropout branch supplies most of the spread
(branch SSR $0.97\to0.92$ from day~1 to day~10), while the encoder branch
supplies a shrinking fraction (branch SSR $0.37\to0.20$; the separately finetuned
chain behaves similarly, Supplementary Fig.~\ref{fig:supp-variants}c). Averaged
over variables and days~1 to 10 the SSR is $0.96$, and per variable it ranges
from $0.97$ (V10, MSLP) down to $0.92$ (T2M; Supplementary
Fig.~\ref{fig:supp-ssr}). The mass fields, which have the weakest analysis
spread, are close to calibrated from day~1 on, so the analysis deficit is not
carried through the rollout. We note that no spread inflation or statistical
post-processing is applied at any stage, which means that the calibration in
Figure~\ref{fig:skill}d is produced by the probabilistic finetuning alone.

\subsection*{Station forecasts are at least as good as Aardvark Weather}

\begin{figure}[!htb]
	\centering
	\includegraphics[width=\linewidth]{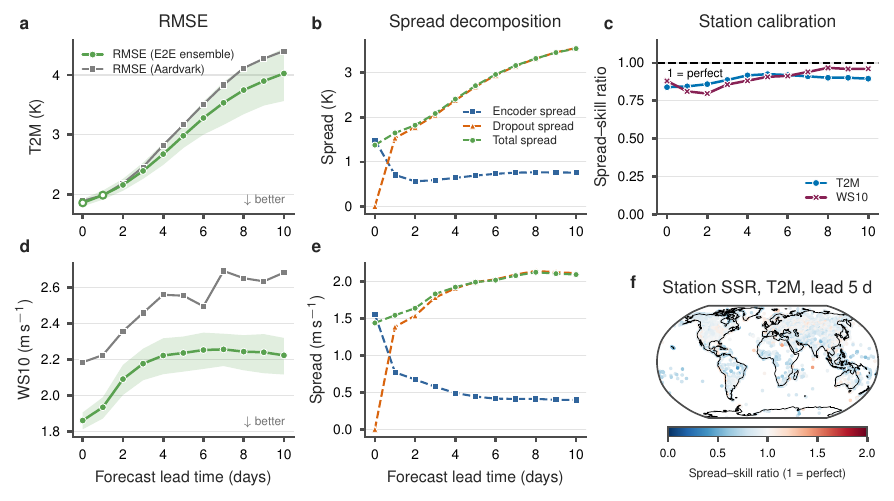}
	\caption{Verification at HadISD surface stations for T2M and WS10, for
		the station-objective end-to-end finetuned E2E ensemble (one finetune
		per variable; Supplementary Note~\ref{note:joint}).
		\textbf{a},\textbf{d}, Ensemble-mean RMSE of the E2E ensemble (green
		circles) and RMSE of the Aardvark forecast (grey squares) for T2M
		(\textbf{a}) and WS10 (\textbf{d}).
		\textbf{b},\textbf{e}, Decomposition of the station ensemble spread
		into its encoder (noise) and dropout branches, together with the total
		spread. Variances add (Eq.~\eqref{eq:ltv}), spreads do not: the total
		spread is the square root of the summed branch variances, so it stays
		close to the dropout spread. \textbf{c}, Station spread--skill ratio
		(SSR) for both variables; the dashed line marks perfect calibration.
		\textbf{f}, Per-station SSR for T2M at a lead time of 5 days.}
	\label{fig:stations}
\end{figure}

The picture carries over to the ground station forecast.\linelabel{rev:R2-5} At
HadISD surface stations, where forecasts are verified for T2M and 10-m wind
speed (WS10), we show the end-to-end finetuned chain, in which encoder,
processor and station decoder are finetuned jointly against the station CRPS
with both stochastic sources active (Supplementary Note~\ref{note:joint}; one
finetune per variable). The deterministic reference is the modular
(non-end-to-end-finetuned) Aardvark configuration. The ensemble mean is at least as accurate as
the deterministic forecast at every lead (Figure~\ref{fig:stations}a,d): for T2M
its RMSE matches Aardvark at analysis time and day~1 and is significantly lower
from day~2 on, by $5$--$9\,\%$ at days~4 to 10; for WS10 it is $10$--$17\,\%$
lower at every lead, significantly so throughout. Evaluated by its decision
value (REV), the station ensemble is
significantly better than the deterministic model at every lead time for T2M:
its maximum REV exceeds Aardvark's by $0.20$--$0.32$ in both tails and at every
lead (Supplementary Note~\ref{note:value}). For WS10 the gain is confined to
the first days: $+0.10$ in both tails at analysis time and day~1, shrinking to
$+0.02$--$0.04$ from day~5 as both forecasts lose their value for station wind
extremes. Calibration is slightly less favorable than on the
grid: the E2E ensemble is close to calibrated but mildly under-dispersive at all
leads (Figure~\ref{fig:stations}c; SSR $0.84$--$0.93$ for T2M and
$0.80$--$0.97$ for WS10). The decomposition of the station spread
(Figure~\ref{fig:stations}b,e) mirrors the gridded attribution: the dropout
branch dominates and grows along the rollout while the encoder branch stays
\linelabel{rev:R1-m4}approximately flat with lead time, holding at most $17\,\%$
(T2M) and $24\,\%$ (WS10) of the station variance from day~1 onward. The
per-station map (Figure~\ref{fig:stations}f) shows the under-dispersion to be
near-global, \linelabel{rev:R1-m5}especially over land, rather than driven by a
few regions. Resolved by region of the Aardvark evaluation protocol
(Supplementary Note~\ref{note:regional}), the ensemble has more decision
value for T2M at every lead in each region, including the sparse West African
and Pacific station sets, whereas for WS10 it degrades to Aardvark value beyond the
first days. The ensemble mean is as accurate as Aardvark or better in the two
dense networks, while the sparse sets are more mixed: over the Pacific it is
less accurate than Aardvark for T2M at days~2 to 4 and for WS10 at most leads
to day~6. The verification-rank histograms\linelabel{rev:R2-7} (Supplementary Fig.~\ref{fig:supp-rank-hist})
confirm the mild under-dispersion.

\subsection*{The encoder branch responds to the observations:
	denying an observing system raises it by up to 84\%}

\begin{figure}[!htb]
	\centering
	\includegraphics[width=\linewidth]{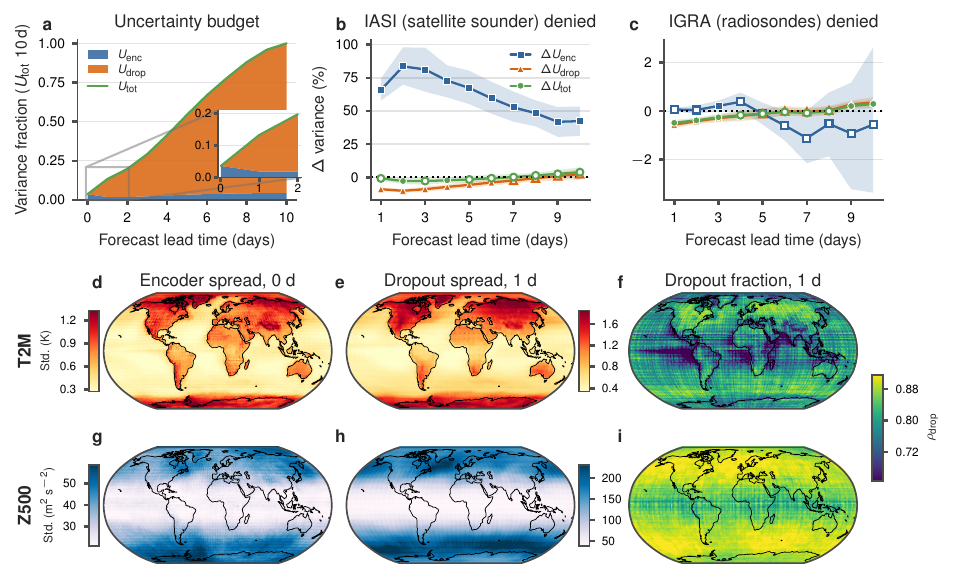}
	\caption{Ensemble variance attributed to the encoder and dropout
		branches across lead times, in space,
		and under observation denial.
		\textbf{a}, Gridded uncertainty budget: the encoder and dropout
		variance components (Eq.~\eqref{eq:ltv}), normalized per variable by
		that variable's total variance at day~10 and averaged over the six
		headline variables, stacked so that they sum to the total (green
		line). At lead 0 (the encoder analysis) all variance lies on the
		encoder branch by
		construction; the inset magnifies leads 0--2.
		\textbf{b},\textbf{c}, Observing-system experiment
		(OSE) cross-check for the most and the least influential streams:
		relative change
		of the encoder ($U_{\mathrm{enc}}$), dropout
		($U_{\mathrm{drop}}$), and total ($U_{\mathrm{tot}}$) variance
		components when the IASI sounder
		(\textbf{b}) or the IGRA radiosonde profiles (\textbf{c}) are
		withheld at encoding time, averaged over all variables, as a function
		of lead time. \textbf{d},\textbf{g}, Encoder ensemble
		spread at analysis time (lead 0), where all spread comes from
		the encoder branch, for T2M (\textbf{d}) and Z500
		(\textbf{g}). \textbf{e},\textbf{h}, Dropout ensemble spread
		at lead time 1 day; note the separate color scales, the encoder
		branch is substantially weaker than the day-1 dropout spread.
		\textbf{f},\textbf{i}, The dropout variance
		fraction $\rho_{\mathrm{drop}}$ implied by the two branch variance
		fields at lead time 1 day. The rectilinear texture in
		\textbf{f},\textbf{i} reflects the difference in patch sizes of the
		encoder and processor ViTs.}
	\label{fig:decomposition}
\end{figure}

\linelabel{rev3:C2-a}To examine whether the stochastic encoder and processor capture distinct sources
of uncertainty, we decompose the total forecast ensemble variance into its
encoder and dropout components (Figure~\ref{fig:decomposition}). The
decomposition behaves consistently with the expected roles of the encoder and dropout branches as observation and model uncertainty, respectively
(Figure~\ref{fig:decomposition}a): the total variance grows from day~1 to day~10
by a factor of $4.5$ (T2M) to $27$ (Z500) while the encoder component grows far less,
so its share falls from all the variance at the analysis (lead~0) to $0.13$ by day~1 and
$0.04$ by day~10. The dropout variance fraction rises correspondingly from $0.87$ to
$0.96$ in the variable mean, matching the NWP expectation that model uncertainty
accumulates along the rollout. The spatial fingerprints refine this reading
(Figure~\ref{fig:decomposition}d--i): at analysis time the encoder spread is
largest in regions that are generally associated with greater uncertainty, such
as high-latitude land, the sea-ice margins, steep orography for T2M and
\linelabel{rev:R1-m6}high latitudes for
Z500\cite{sandu2013stable,wallace1989sstwind}. The dropout spread at day~1
concentrates over high-latitude land and sea ice for T2M and
in the extratropical storm tracks for Z500. The dropout fraction for T2M dips
over the equatorial eastern Pacific, possibly reflecting the influence of the
slowly varying ocean surface temperature on boundary-layer
variability\cite{barsugli1998coupling}, which would damp the growth of processor
perturbations more than the encoder, and over steep orography and the ice
sheets, where the encoder spread (Figure~\ref{fig:decomposition}d) is large
relative to the dropout spread, as shown in the ratio panel
Figure~\ref{fig:decomposition}f. \linelabel{rev:R1-m7} In these regions
ensemble data assimilation places its largest analysis
spread\cite{bonavita2012eda}. For Z500 it dips throughout the tropics,
consistent with the weak tropical error growth and low geopotential variance
there\cite{zagar2017limits}.

\linelabel{rev3:C2-b}The reading of the encoder branch as
observation-driven uncertainty is consistent with a mechanistic test in the
style of observing-system experiments (OSEs, see \hyperref[sec:ose]{Methods}):
withholding a single observation stream at encoding time and re-running the
ensemble moves mainly the encoder-attributed component
(Figure~\ref{fig:decomposition}b,c). Denying the IASI sounder, which is the most
influential stream, raises $U_{\mathrm{enc}}$ by $66$--$84\,\%$ at days~1--4
(variable mean), decaying to $+42\,\%$ by day~10 as the denied information is
forgotten by the forecast dynamics, while $U_{\mathrm{drop}}$ responds by at
most about $10\,\%$ (largest response $-10\,\%$, at day~2).
\linelabel{rev:R2-8}The total variance barely moves ($-3$ to $+4\,\%$): the
encoder component holds only $13\,\%$ of the total at day~1 and $4\,\%$ at
day~10, so its large relative response is a small absolute one, offset at short
leads by the slight decrease of the dropout component. Denying the least
influential stream (the IGRA radiosondes) leaves every component within about
$1\,\%$ of baseline.
The denial response is larger for the more influential stream and does not
inflate the dropout axis in either experiment. \linelabel{rev3:C2-c}With this evidence we adopt, as an
approximation, the interpretation the two branches were designed for: the
encoder branch as an approximately aleatoric (observation-driven) contribution
and the dropout branch as an approximately epistemic (model-driven)
contribution, $U_{\mathrm{enc}}\approx U_{\mathrm{alea}}$ and
$U_{\mathrm{drop}}\approx U_{\mathrm{epis}}$. This is an attribution aligned with
the components of the model, not a recovery of the underlying uncertainty types
(see \hyperref[sec:vardecomp]{Methods}). 
The approximation is one-sided: the encoder share is a lower bound on the
observation contribution and the dropout share an upper bound on the model
contribution (see \hyperref[sec:vardecomp]{Methods}). \linelabel{rev:R1-M6}A
control experiment on the separately finetuned chain sizes the lower bound: with
the processor finetuned on deterministic rather than stochastic analyses, so
that its dropout branch cannot absorb analysis uncertainty, the day-1 encoder
share of that chain rises from $0.14$ to $0.25$ due to a smaller dropout spread,
and the two agree within $0.02$ from day~5, at the cost of a less skillful
ensemble at short lead times. This control experiment indicates that part of the
dropout branch's spread in the E2E ensemble compensates for the encoder's
analysis uncertainty, but the effect is confined to days~1 to 3 and does not
change the qualitative conclusions. The encoder branch responsiveness has a practical use.
Where a classical denial study requires a full NWP or adjoint assimilation
system \cite{bouttier2001ose,langland2004fsoi}, in the E2E ensemble withholding
a stream amounts to re-encoding and re-running the ensemble, so the contribution
of each observing system to forecast confidence can be quantified. This can be
useful for planning future observing networks, e.g.\ to identify gaps in
existing coverage.

Taken together, these results show that a deterministic end-to-end forecast
system can be uplifted to a stochastic one at a gain to the forecast qualities
it already had, starting at the analysis itself: the stochastically finetuned
encoder produces analysis states $14\,\%$ better in ensemble-mean RMSE than the
deterministic system's, and the advantage persists at every forecast lead
(Figure~\ref{fig:skill}a). Judged by its decision value, the ensemble's maximum
relative economic value exceeds the deterministic forecast's by $0.15$--$0.31$
(variable mean) at every lead time, significantly for every variable, tail and
lead time, retaining $0.32$ of a perfect forecast's value at day~10 where the
deterministic forecast sinks to $0.07$, and is close to calibrated on the grid
throughout the medium range (variable-mean spread--skill ratio $0.96$ over
days~1--10) without any post-hoc spread adjustment, while station forecasts
match or beat the deterministic baseline in RMSE at
every lead (tied at the analysis and day~1 and significantly better from day~2
for T2M, significantly better at every lead for WS10, Figure~\ref{fig:stations}a,d).
The two sources of stochasticity remain
interpretable: \linelabel{rev3:C2-d}the encoder branch approximates the
observation-driven (aleatoric) contribution and the dropout branch the
model-driven (epistemic) contribution, whose share of the total variance grows
from $0.87$ at day~1 to $0.96$ at day~10. The results from the observation-denial experiment are
consistent with this component-aligned attribution
(Figure~\ref{fig:decomposition}).

\section*{Discussion}\phantomsection\label{sec:discussion}

\linelabel{rev:R1-m8}Uncertainty decomposition makes it possible, in principle,
to determine from a forecast whether its spread originates in the initial state, which
the observations constrain incompletely, or in the forecast dynamics. These two
readings call for different remedies, new or better-placed observations in the
first case, more training data, better architectures or larger ensembles in the
second, and an ensemble driven by a single noise
source\cite{price2025gencast,langAIFSCRPSEnsembleForecasting2026} has no way to
distinguish them. \linelabel{rev:R2-9}A single source can be tuned to a
calibrated total spread, but it answers only how uncertain a forecast is, not
why. \linelabel{rev3:C2-f}The disentanglement is what turns the ensemble into
a diagnostic of the observing system, as the denial experiment illustrates. Obtaining the distinction here
required neither a new model nor a full generative treatment: finetuning two
noise sources on top of a pretrained deterministic system, the nested-ensemble
uplift proposed here, is sufficient.

Within this frame, the three questions posed in the introduction can be answered
directly. On \emph{disentanglement}, attaching one random source to each
component makes the uncertainty decomposable: the system states, at inference
time, how much of its spread enters through the encoder and how much through the
processor, a separation NWP obtains from its ensemble
data-assimilation machinery\cite{isaksen2010eda,bonavita2012eda}. On
\emph{cost}, probabilistic output and deterministic skill are not in tension in
observation-to-forecast models: the relatively cheap stochastic finetuning
improves the mean forecast significantly at every variable and lead time,
beginning with the analysis state itself. On the \emph{explanation} of forecast
uncertainty, the attribution departs from an intuition inherited from NWP
ensembles\cite{molteni1996eps,leutbecher2008ensemble}. Its direction matches:
model uncertainty grows in relative importance along the rollout. Its level does
not. Even at day~1, where classical ensembles are dominated by initial-condition
uncertainty, about a seventh of the predictive variance is attributed to the
encoder branch and the rest to the dropout branch. Independently of the aleatoric/epistemic labels, the robust reading is
a pair of bounds: the encoder share is a lower bound on the true observation
(aleatoric) contribution, and the dropout share an upper bound on pure model
uncertainty. \linelabel{rev:R1-M6_2} The control experiment with the processor
puts the day-1 encoder share at a quarter rather than a seventh, with the bounds
converging from day~5. Three non-exclusive mechanisms could produce this
imbalance, and each merits its own follow-up investigation. The learned encoder
noise may under-represent observation and assimilation uncertainty, consistent
with the under-dispersion of the analysis ensemble ($\mathrm{SSR}=0.60$); the
EDA comparison proposed below is the direct test. The conditioning order of the
variance decomposition folds the encoder--processor interaction into the dropout
term; a crossed design that freezes the dropout masks while varying the encoder
draws would measure that interaction separately. And finally, MC dropout with
masks redrawn at every step acts as a stochastic model perturbation that
accumulates along the rollout, closer to stochastic
physics\cite{palmer2009stochastic,leutbecher2017stochastic} than to a fixed
posterior draw; rollouts holding a single mask per member would isolate the
accumulation.

Two findings transfer directly to the design of future stochastic end-to-end
systems. First, where noise enters matters: distributing the same learned
perturbation across all encoder blocks through multiplicative conditioning
yields better analyses and forecasts, at less raw spread, than a single additive
injection (Supplementary Note~\ref{note:variants}). \linelabel{rev:R1-M8_2}
Second, observation denial gives a cheap, direct test of the uncertainty
attribution: withholding a stream must raise the observation branch and leave
the model branch unchanged, and the ensemble can be checked against this at
negligible cost. Where a classical observing-system experiment re-runs a full
assimilation--forecast cycle over months of dates\cite{bouttier2001ose},
withholding a stream here amounts to re-encoding and re-running the ensemble, a
few hours on a single GPU (Fig.~\ref{fig:decomposition}b,c).

The main limitations of this work are structural, and each points to its own
remedy. The decoder is deterministic, so no uncertainty is added at the
downscaling stage; \linelabel{rev3:E-a}the station-level under-dispersion points to a third
stochastic branch at the decoder. The dropout posterior is a restricted one: the
mask rate is a fixed hyperparameter ($p=0.05$) rather than a learned quantity,
so the dropout spread is calibrated only indirectly, through the adaptation of
the processor weights under the fair-CRPS objective. \linelabel{rev:R1-M4}A
learned rate, as in concrete dropout\cite{gal2017concrete}, would let the
dropout spread respond to the amount of training data, which a fixed rate
cannot, and deep ensembles\cite{lakshminarayanan2017deepensembles} of processors
would provide a stronger epistemic reference at a higher training cost.
\linelabel{rev:R2-2-disc}The assignment of one mechanism per component is a
design choice: swapping the two, or placing both in both
components, would separate what the branch behavior owes to the mechanism from
what it owes to the component, and is a direct follow-up to this study.
Verification covers a single held-out year, one
backbone\cite{allen2025aardvark}, and modest ensemble sizes, so the
non-negativity floor of the encoder-variance estimator in Eq.~\eqref{eq:anova}
is occasionally active at long leads. The backbone itself, finally, is
deliberately lean: a $1.5^\circ$ grid and under a terabyte of preprocessed
observations from nine streams, a small fraction of the operationally
assimilated observing system. Part of the remaining gap to IFS ENS, largest at
the short leads where resolution and observation coverage matter most, is
therefore attributable to the backbone rather than to the probabilistic
machinery, and the skill and calibration reported here should be read as a lower
bound on what the approach can deliver at higher resolution and with more
observation data. The aleatoric/epistemic reading of the encoder and dropout
branches likewise remains an approximation, aligned with the components of the
model rather than with recovered ground-truth uncertainty
types\cite{huellermeier2021aleatoric}.

Two extensions follow naturally. First, the learned encoder spread invites a
systematic comparison against the ERA5 Ensemble of Data Assimilations
(EDA)\cite{isaksen2010eda,bonavita2012eda}, the closest physically grounded
reference for assimilation uncertainty and the direct test of the
under-representation reading above, treating the EDA's documented
under-dispersion as a qualitative reference rather than ground truth. Second,
because the encoder branch is attached to the observation interface, the denial
experiment of Figure~\ref{fig:decomposition}b,c extends from attribution to
observing-network design: sweeping the ingested observing systems in turn to ask
not only which streams the analysis depends on, but which additions,
degradations, or losses would most change forecast confidence. Questions of the
form ``which stream's loss would most inflate day-3 forecast uncertainty?'' are
the core transaction of an observation-driven digital twin of the atmosphere,
and \linelabel{rev3:E-c}the decomposed spread can address them here at inference cost. The decomposition
also opens the door to auditing corrupted or adversarially perturbed observation
inputs, although this would require further study.

\section*{Methods}\phantomsection\label{sec:methods}

\subsection*{End-to-end weather forecast as a stochastic map}\phantomsection\label{sec:stochmap}

Figure~\ref{fig:method} summarizes the proposed framework. The deterministic
Aardvark system\cite{allen2025aardvark} is the composition of three learned
stages, $f_\theta = D_\theta \circ P_\theta^{(\tau)} \circ E_\theta$: the
encoder $E_\theta$ maps the heterogeneous observations $\mathbf{x}$ of an
assimilation window to an initial state $\mathbf{z}_0$ (internalizing data
assimilation; $30.7$ million parameters), the processor $P_\theta$ advances the
state autoregressively over $\tau$ daily steps ($\mathbf{z}_\tau$; $54.0$
million parameters per step), and the decoder $D_\theta$ maps the forecast state
to the target stations ($21.1$ million parameters per variable and lead time).
The encoder and processor are Vision Transformers\cite{dosovitskiy2021vit} and
the decoder a U-Net\cite{ronneberger2015unet}. As published, $f_\theta$ issues a
single forecast with no quantified uncertainty. We uplift it to a stochastic map
$g$ by attaching one random source to each of the encoder and processor
(Fig.~\ref{fig:method}a):
\begin{equation}
	\hat{\mathbf{y}}_\tau \;=\; g(\mathbf{x};\,\mathbf{a},\boldsymbol{\omega})
	\;=\;
	D_\theta\!\Big(P_{\boldsymbol{\omega}}^{(\tau)}\big(E_\theta(\mathbf{x};\mathbf{a})\big)\Big),
	\label{eq:functional}
\end{equation}
with $\hat{\mathbf{y}}_\tau$ the forecast fields at lead $\tau$, and the
encoder-noise draw $\mathbf{a}$ (the observation-side source) and the rollout
dropout masks $\boldsymbol{\omega}$ (the model-side source) sampled
independently. The learned weights $\theta$, fixed at inference, are suppressed
in $g$; the arguments after a semicolon are random draws. Throughout the
Methods, bold lowercase symbols denote fields or vectors and italic letters
scalars, superscripts $(i,j)$ index the encoder-noise draw and the dropout
member and hats (e.g. $\hat{U}_{\mathrm{tot}}$) denote finite-ensemble
estimates. Supplementary Table~\ref{tab:supp-notation} collects all symbols used
throughout the paper.

\emph{Encoder branch.} The encoder is made stochastic through learned,
input-dependent Gaussian noise, $E_\theta(\mathbf{x};\mathbf{a}) =
	\boldsymbol{\mu}_\theta(\mathbf{x}) +
	\boldsymbol{\sigma}_\theta(\mathbf{x})\odot\mathbf{a}$, where
$\boldsymbol{\mu}_\theta(\mathbf{x})$ is the deterministic analysis of the
pretrained encoder, $\boldsymbol{\sigma}_\theta(\mathbf{x})$ a learned,
input-dependent noise amplitude,
$\mathbf{a}\sim\mathcal{N}(\mathbf{0},\mathbf{I})$ the noise draw, and $\odot$
the element-wise product. This follows the heteroscedastic aleatoric
construction of Kendall and Gal\cite{kendall2017uncertainties} in spirit, placed
at the assimilation interface and calibrated through the CRPS on sampled members
rather than through a likelihood, so that observation-sparse or low-quality
regions can be assigned larger latent noise. This noise injection approach is
similar to the one used in AIFS-CRPS\cite{langAIFSCRPSEnsembleForecasting2026}.
Concretely, $\mathbf{a}$ is drawn per patch, mapped by a small learned MLP to a
per-patch embedding, and injected into the pretrained encoder backbone in one of
two variants (Supplementary Note~\ref{note:variants}): \emph{embedding
	injection} adds it once to the token embeddings before the first transformer
block ($+0.3$ million parameters), while \emph{norm
	conditioning} re-injects the same realization at every block through
conditional LayerNorms in the style of feature-wise linear modulation
(FiLM)\cite{perez2018film}, whose zero-initialized projection leaves the
pretrained encoder unchanged at initialization ($+8.7$ million parameters). The
results in this paper use the norm-conditioning variant, which, compared as
separately finetuned chains, achieves a $4$--$5\,\%$ lower variable-mean CRPS
than embedding injection at short and medium leads at equal ensemble size, while
leaving the uncertainty attribution essentially unchanged (Supplementary
Fig.~\ref{fig:supp-variants}). The amplitude
$\boldsymbol{\sigma}_\theta(\mathbf{x})$ and the injection pathway receive no
direct supervision: they are learned by finetuning the encoder with a fair CRPS
objective over $M$ noise draws against ERA5 analyses (see
\hyperref[sec:training]{Data and training}). \linelabel{rev:R1-m9}CRPS is a
proper score, so minimizing it rewards calibration and sharpness jointly, but it
does not enforce calibration \cite{wilks2018enforcing}.

\emph{Dropout branch.} The processor weights are treated as random via MC
dropout, recast as approximate Bayesian inference\cite{gal2016dropout}: dropout
($p=0.05$) stays active at inference, and one realization $\boldsymbol{\omega}$
denotes the collection of masks encountered over the $\tau$-step rollout,
redrawn at every lead time and for every ensemble member; the subscript in
$P_{\boldsymbol{\omega}}^{(\tau)}$ records that these masks act on the processor
weights at each lead time. The rate $p=0.05$ was selected in a sweep on the
Aardvark processor (inference-time dropout at $p\in\{0.05,0.1,0.2\}$ over
ERA5-initialized rollouts) as the optimal value as measured by the spread--skill
ratio and ensemble mean RMSE. \linelabel{rev:R1-M5}The processor is finetuned
with dropout active ($p=0.05$), taking as input the analyses
produced by the stochastic encoder and minimizing the fair CRPS of its dropout
ensemble against ERA5. The finetuned weights, and hence the spread that the
dropout masks generate, are therefore matched to the stochastic encoder the
processor is composed with. The decoder is kept deterministic, so each random
source is attached to exactly one component and the split below is a strict
two-way split.\linelabel{rev:R2-2-design} This assignment of different
mechanisms to the encoder and processor intends to mirror operational practice,
with perturbed analyses for initial-condition uncertainty and stochasticity
inside the forecast model for model uncertainty.

\subsection*{Variance decomposition and nested-ensemble uncertainty estimation}\phantomsection\label{sec:vardecomp}

Let $y$ denote a single scalar component of the forecast
$\hat{\mathbf{y}}_\tau$. This is one variable at one grid point and lead time;
all quantities below are computed element-wise over these axes. Given the
observations $\mathbf{x}$, the only randomness in $y$ comes from the two
independent sources $\mathbf{a}\sim\mathcal{N}(\mathbf{0},\mathbf{I})$ and
$\boldsymbol{\omega}\sim q(\boldsymbol{\omega})$, the dropout-mask
distribution\cite{gal2016dropout}; every expectation and variance below is taken
over this product space, with subscripts indicating the source being integrated
out. Conditioning the predictive variance on the encoder-noise draw and applying
the law of total variance gives the exact split:
\begin{equation}
	\underbrace{\operatorname{Var}[\,y\mid\mathbf{x}\,]}_{U_{\mathrm{tot}}}
	\;=\;
	\underbrace{\operatorname{Var}_{\mathbf{a}}\!\big[\mathbb{E}_{\boldsymbol{\omega}}[\,y\mid\mathbf{x},\mathbf{a}\,]\big]}_{U_{\mathrm{enc}}}
	\;+\;
	\underbrace{\mathbb{E}_{\mathbf{a}}\!\big[\operatorname{Var}_{\boldsymbol{\omega}}[\,y\mid\mathbf{x},\mathbf{a}\,]\big]}_{U_{\mathrm{drop}}},
	\label{eq:ltv}
\end{equation}
where $U_{\mathrm{tot}}$ is the total predictive variance and $U_{\mathrm{enc}}$
and $U_{\mathrm{drop}}$ are its encoder (between-draw) and dropout (within-draw)
components.

In the aleatoric/epistemic sense of the deep learning uncertainty quantification
literature the two components are read as $U_{\mathrm{enc}}\approx
	U_{\mathrm{alea}}$ (observation-driven) and $U_{\mathrm{drop}}\approx
	U_{\mathrm{epis}}$ (model-driven): this is the variance form of the
decomposition of Depeweg et al.\cite{depeweg2018decomposition} with the encoder
noise playing the role of the aleatoric latent and the processor weights the
epistemic one. Whether an uncertainty is aleatoric or epistemic depends on what
is held fixed\cite{derkiureghian2009aleatory}: here the observing system and the
processor architecture. The reading above is meant in this relative sense:
the split identifies which uncertainties would respond to further information.
Because $g$ is nonlinear, the
$\mathbf{a}$--$\boldsymbol{\omega}$ interaction is absorbed into the dropout
term by the conditioning order (we condition on $\mathbf{a}$ because it keeps
$U_{\mathrm{enc}}$ \linelabel{rev3:C2-g}as the effect of the encoder noise alone), so $U_{\mathrm{drop}}$ is
an upper bound on the dropout contribution. \linelabel{rev:R1-M7-methods}The
identification encoder$\to$aleatoric, dropout$\to$epistemic is an approximation.
The split is an attribution to the components of the model, not a recovery of
ground-truth uncertainty
types\cite{huellermeier2021aleatoric,valdenegro2022deeper}; for this reason we
name the two terms after the components, encoder and dropout, and test the
approximation with the observation-denial experiment below.

The expectations in Eq.~\eqref{eq:ltv} are estimated with a nested ensemble
(Fig.~\ref{fig:method}b): $M$ outer encoder-noise draws $\mathbf{a}^{(i)}$, each
frozen and rolled out under $N$ inner MC-dropout realizations, giving $MN$
members $\hat{y}^{(i,j)}$ with per-draw means $\bar{y}^{(i)} =
	\tfrac{1}{N}\sum_{j}\hat{y}^{(i,j)}$ and grand mean $\bar{y}$. The within- and
between-group mean squares of a one-way random-effects ANOVA:
\begin{equation}
	\mathrm{MS}_{\mathrm{W}}
	= \frac{1}{M(N-1)}\sum_{i=1}^{M}\sum_{j=1}^{N}
	\big(\hat{y}^{(i,j)}-\bar{y}^{(i)}\big)^{2},
	\qquad
	\mathrm{MS}_{\mathrm{B}}
	= \frac{N}{M-1}\sum_{i=1}^{M}\big(\bar{y}^{(i)}-\bar{y}\big)^{2},
	\label{eq:ms}
\end{equation}
$\mathrm{MS_W}$ measures how the $N$ dropout members scatter around their own
group's mean. This spread is attributable to the processor alone, since the
encoder draw is fixed within a group, while $\mathrm{MS_B}$ measures how the $M$
group means scatter around the grand mean, reflecting the effect of changing the
encoder draw. The estimators of the two branches, unbiased under the
random-effects model, are:
\begin{equation}
	\widehat{U}_{\mathrm{drop}} = \mathrm{MS}_{\mathrm{W}},
	\qquad
	\widehat{U}_{\mathrm{enc}} =
	\max\!\big(0,\ (\mathrm{MS}_{\mathrm{B}}-\mathrm{MS}_{\mathrm{W}})/N\big).
	\label{eq:anova}
\end{equation}
The $\mathrm{MS}_{\mathrm{W}}/N$ subtraction removes the finite-$N$ sampling
noise that each group mean inherits from the dropout draws; without it, dropout
variance leaks into the encoder estimate and systematically overstates the
observation branch. One pass of the nested ensemble yields the full spatial
fields behind Figure~\ref{fig:decomposition}. The attribution ratio reported
there is
$\rho_{\mathrm{drop}}=\widehat{U}_{\mathrm{drop}}/(\widehat{U}_{\mathrm{enc}}+\widehat{U}_{\mathrm{drop}})$.
\linelabel{rev:R2-6}Every decomposition shown in this paper, gridded,
station-level and under observation denial, is computed from this one nested
ensemble; no ensemble with a single perturbed source is used.

\subsection*{Observation-denial cross-check}\phantomsection\label{sec:ose}

\linelabel{rev:R1-M8}Observing-system experiments withhold one observation type
and measure how much the analysis degrades\cite{bouttier2001ose}. We use the
same intervention to test the variance decomposition. Withholding an observation
stream removes information from the encoder only; the processor and its dropout
masks are untouched. If the attribution holds, the
between-encoder (observation) variance must increase while the dropout (model)
variance stays unchanged. The decomposition can therefore fail this test in a
specific way: an increase of the dropout variance under denial would show that
the dropout branch also responds to observation uncertainty.
Figure~\ref{fig:decomposition}b,c reports the outcome. \linelabel{rev:R2-3} A
stream $\mathbf{x}_s$ is withheld by zeroing its gridded encoder embedding. The
density-aware SetConv observation interface already produces a zero embedding
wherever an observation stream has no observations (in satellite swath gaps, or
no stations present in that grid cell), so a zeroed embedding is equivalent to
no data available for a certain stream. Baseline and denial configurations share
per-initialization random seeds (common random numbers): the conditioning-noise
and dropout-mask draws are identical, so the per-initialization changes are
paired. We perform this cross-check across $24$ initializations spanning the
test year, at the full nested configuration of the main evaluation ($M{=}7$
encoder draws, $N{=}7$ dropout rollouts each).

\subsection*{Data and training}\phantomsection\label{sec:training}

All experiments use the publicly released Aardvark data
configuration\cite{allen2025aardvark}, distributed as a preprocessed dataset on
Hugging Face\cite{anna_vaughan_2025}. The observational input comprises the
original Aardvark system's heterogeneous streams: HadISD surface station
reports\cite{dunn2012hadisd} ($23$\,GB), ICOADS marine reports ($11$\,GB), IGRA
radiosonde profiles ($4$\,GB), AMSU-A ($69$\,GB), AMSU-B ($64$\,GB), HIRS
($139$\,GB) and IASI ($296$\,GB) sounder radiances, ASCAT scatterometer winds
($91$\,GB), and GridSat geostationary satellite imagery ($37$\,GB): about
$0.73$\,TB of observations in total over the full preprocessed archive. ERA5
\cite{hersbach2020era5} ($156$\,GB) additionally provides the training targets
for the encoder and processor (24 surface and pressure-level variables on the
$1.5^\circ$ global grid) and the gridded verification ground truth, while HadISD
provides the station-level verification targets (T2M and WS10). Following the
original paper's protocol\cite{allen2025aardvark}, 2007--2017 is used for
training, 2019 for validation and checkpoint selection, and 2018 is held out for
testing.

All components are warm-started from the published deterministic Aardvark
weights\cite{allen2025aardvark} and finetuned separately and then jointly, with a fair
(finite-ensemble-corrected) CRPS
objective\cite{hersbach2000crps,leutbecher2008ensemble}. The noise-conditioned
encoder is finetuned against ERA5 analyses at 00\,UTC with $M=7$ noise members
per step (AdamW, learning rate $3\times10^{-5}$, weight decay $10^{-5}$, cosine
schedule, 20 epochs, batch size 4 per GPU). \linelabel{rev:R2-4-members}$M=7$ is
the smallest member count tested whose analysis ensemble mean improves on the
deterministic Aardvark analysis for every headline variable ($M=4$ does not),
and it is kept for all stages, also so that the $7\times7$ inference ensemble is comparable with the 50-member IFS ENS in size.
The processor is then finetuned, with gradient
checkpointing, on the stochastic encoder's initial conditions,
\linelabel{rev:R2-4}generated once per initialization with a single noise draw
and stored, with dropout active ($p=0.05$) and $7$ dropout members per step on
that analysis (AdamW, learning rate $10^{-4}$ annealed to $10^{-6}$, weight
decay $10^{-5}$), sequentially for lead times 1--10, each lead initialized from
the previous one and trained on its output with 20 epochs at lead~1 and 12
epochs per subsequent lead. As in the deterministic Aardvark
protocol\cite{allen2025aardvark}, the processor is trained in the marginal: at
each step it predicts the normalized difference to the previous daily state,
which is added back to that state to advance the rollout, rather than predicting
the full atmospheric field directly. In this separately finetuned chain the
station decoders are those of the published Aardvark release, kept deterministic
and frozen; the chain is then finetuned end-to-end against the gridded or the
station target (Supplementary Note~\ref{note:joint}), the station objective
retraining the station decoders jointly with encoder and processor. The results
of the main text are those of the end-to-end finetuned chains. The component
stage, with 4-GPU data parallelism on NVIDIA A100 (80\,GB) nodes, takes
${\approx}4.7$\,h wall-clock for the encoder finetune and ${\approx}11.5$\,h for
the sequential lead-1--10 processor finetune: about 65 A100-hours for one
complete probabilistic chain. The joint stage (Supplementary
Note~\ref{note:joint}) adds roughly one further day on a single node of NVIDIA
B200 (183\,GB) GPUs per objective: 27\,h on 4 GPUs for the gridded objective and
12--13\,h on 8 GPUs per station variable, about 100 B200-hours each. In total,
the system is obtained from the pretrained deterministic chain in under two days
of wall-clock time on one multi-GPU node at a time. No data-loading or other
optimizations were performed during training, so these figures are upper bounds
on training duration. The deterministic Aardvark baseline in
Figures~\ref{fig:skill}a,b and~\ref{fig:stations}a,d is the published
encoder--processor--decoder chain evaluated identically. Note that the published
Aardvark station results additionally use an end-to-end finetune of the full
system; since weights for that configuration were released for a single lead
time only and we were unable to reproduce the published station scores with
them, we use the modular (non-end-to-end-finetuned) release as the deterministic
reference at all lead times.

\subsection*{Experimental setup and verification metrics}\phantomsection\label{sec:metrics}

At inference, ensembles are generated by crossing $M$ encoder-noise draws with
$N$ processor dropout masks (here $M=7$, $N=7$, i.e.\ 49 members). This
configuration matches the member counts used during finetuning, yields a total
ensemble comparable in size to the 50-member IFS ENS reference, and balances the
sampling precision of the two branch estimators in Eq.~\eqref{eq:anova}.
Forecasts are verified against ERA5 over the $337$ initializations of the
held-out year 2018 on the $1.5^\circ$ global grid with latitude-weighted
scores following the WeatherBench~2 protocol\cite{rasp2024weatherbench2}, for
the six headline variables T2M, T850, U10, V10, Z500 and MSLP; IFS ENS
reference scores are computed against the same ERA5 targets. All evaluation is restricted
to 00\,UTC valid time for every model compared. Station forecasts are verified
against HadISD\cite{dunn2012hadisd} T2M and WS10 station observations over the
same initializations, restricted to the station subset used by the original
Aardvark system\cite{allen2025aardvark}.

Probabilistic skill is measured with the fair (finite-ensemble-corrected)
CRPS\cite{hersbach2000crps,leutbecher2008ensemble}.
\linelabel{rev:R2-2-ssr}Calibration is measured with the spread--skill ratio in
the unbiased form of WeatherBench\cite{rasp2024weatherbench2,weatherbenchx2024}.
For IFS ENS the ratio is the unbiased spread--skill ratio published with the
WeatherBench~2 scores. The ensemble-mean RMSE reported elsewhere is the RMSE of
the realized $K$-member mean, without this debiasing. The ratio is computed for
the total ensemble and per branch, where each branch's spread is
$\sqrt{\widehat{U}_{\mathrm{enc}}}$ or $\sqrt{\widehat{U}_{\mathrm{drop}}}$ from
Eq.~\eqref{eq:anova}. Gridded scores are latitude-weighted spatial means;
station scores are means over the verified station population.

\linelabel{rev:R1-M2-metrics}Decision value is measured with the relative
economic value (REV)\cite{richardson2000value} of the cost--loss
model\cite{murphy1977costloss}: a user who pays $C$ to avoid a loss $L$ protects
whenever the forecast probability of the event is at least $C/L$, and REV is the
resulting saving over climatology as a fraction of the saving from a perfect
forecast (0: climatology, 1: perfect). Events are exceedances of the local 10th
and 90th percentiles of the 2007--2017 climatology, the forecast
probability is the member fraction, and the deterministic forecast issues
$p\in\{0,1\}$. These moderate thresholds are chosen to keep the comparison
fair to the deterministic forecast. Figure~\ref{fig:skill}b reports the difference
$\mathrm{REV}_{\mathrm{ens}}-\mathrm{REV}_{\mathrm{det}}$ at the cost--loss
ratio where REV peaks, for ease of
interpretation. Full REV curves are in Supplementary Note~\ref{note:value}.

All statistical inference is carried out on per-initialization score series. For
every model, variable, and lead time, scores are first aggregated over space
(latitude-weighted grid mean or station mean), as in the WeatherBench~2
protocol\cite{rasp2024weatherbench2}, so that each of the $n=337$
initializations contributes exactly one value. Spatial correlation is thereby
absorbed before any test is applied; the only remaining dependence is the
temporal autocorrelation of the series, which every procedure below accounts for
through a common, data-adaptive correlation length $\ell$:
\begin{equation*}
	\ell = \max\!\bigl(n^{1/3},\; 3\,(1+r_1)/(1-r_1)\bigr),
\end{equation*}
where $r_1$ is the lag-1 autocorrelation of the series at hand and $n$ the
number of initializations. Confidence intervals, shown as the shaded bands in
the figures, are 95\% percentile intervals from a circular moving-block
bootstrap ($5\times10^3$ resamples) with block length $\ell$. Nonlinear
aggregates (RMSE, spread--skill ratios, relative changes) are recomputed on each
resample rather than approximated. When two models are compared, both score
series are resampled with the same block indices, preserving the pairing. Significance of paired model differences, shown as filled versus open markers in
the figures, is assessed with the Diebold--Mariano
test\cite{diebold1995comparing} applied to the score-difference series, with
an autocorrelation-robust variance that uses lags up to $\ell$. This is equivalent to the autocorrelation-corrected paired
$t$-test\cite{geer2016significance} that is standard in the verification of
data-driven forecasts\cite{lam2023graphcast}. Markers are filled only where the
difference is significant at the 5\% level. These conventions are used in every
figure: shaded bands are the 95\% confidence intervals above, filled markers
denote a difference from the reference that is significant at the 5\% level
and open markers one that is not, and dashed lines mark the reference (zero
change, parity, or a spread--skill ratio of one).

Large-language-model assistants (Claude Code, Anthropic) were used for code
assistance, manuscript editing and literature search; all AI-assisted output was
reviewed and verified by the authors, who take full responsibility for the
content of this manuscript.

\section*{Data availability}

The finetuned model weights and the evaluation data generated in this study are
available at \href{https://doi.org/10.57967/hf/11853}{https://doi.org/10.57967/hf/11853}. The
training and verification inputs are the publicly released Aardvark
dataset\cite{anna_vaughan_2025}, which includes the HadISD station
records\cite{dunn2012hadisd}. The ERA5 fields\cite{hersbach2020era5} and IFS ENS
used for verification were fetched from WeatherBench
\cite{rasp2024weatherbench2}.

\section*{Code availability}

\linelabel{rev3:C1}The code to reproduce all experiments and figures
is available at \href{https://doi.org/10.5281/zenodo.24534696}{https://doi.org/10.5281/zenodo.24534696}.

\bibliography{lit_v2}

\begin{thebibliography}{10}
\urlstyle{rm}
\expandafter\ifx\csname url\endcsname\relax
  \def\url#1{\texttt{#1}}\fi
\expandafter\ifx\csname urlprefix\endcsname\relax\def\urlprefix{URL }\fi
\expandafter\ifx\csname doiprefix\endcsname\relax\def\doiprefix{DOI: }\fi
\providecommand{\doilink}[1]{\href{https://doi.org/#1}{\nolinkurl{#1}}}
\providecommand{\bibinfo}[2]{#2}
\providecommand{\eprint}[2][]{\url{#2}}

\bibitem{bauer2015quiet}
\bibinfo{author}{Bauer, P.}, \bibinfo{author}{Thorpe, A.} \&
  \bibinfo{author}{Brunet, G.}
\newblock \bibinfo{journal}{\bibinfo{title}{The quiet revolution of numerical
  weather prediction}}.
\newblock {\emph{\JournalTitle{Nature}}} \textbf{\bibinfo{volume}{525}},
  \bibinfo{pages}{47--55}, \doiprefix\doilink{10.1038/nature14956}
  (\bibinfo{year}{2015}).

\bibitem{pathak2022fourcastnet}
\bibinfo{author}{Pathak, J.} \emph{et~al.}
\newblock \bibinfo{journal}{\bibinfo{title}{{FourCastNet}: A global data-driven
  high-resolution weather model using adaptive fourier neural operators}}.
\newblock {\emph{\JournalTitle{arXiv preprint}}}
  \doiprefix\doilink{10.48550/arXiv.2202.11214} (\bibinfo{year}{2022}).

\bibitem{bi2023pangu}
\bibinfo{author}{Bi, K.} \emph{et~al.}
\newblock \bibinfo{journal}{\bibinfo{title}{Accurate medium-range global
  weather forecasting with {3D} neural networks}}.
\newblock {\emph{\JournalTitle{Nature}}} \textbf{\bibinfo{volume}{619}},
  \bibinfo{pages}{533--538}, \doiprefix\doilink{10.1038/s41586-023-06185-3}
  (\bibinfo{year}{2023}).
\newblock \bibinfo{note}{ArXiv:2211.02556}.

\bibitem{lam2023graphcast}
\bibinfo{author}{Lam, R.} \emph{et~al.}
\newblock \bibinfo{journal}{\bibinfo{title}{Learning skillful medium-range
  global weather forecasting}}.
\newblock {\emph{\JournalTitle{Science}}} \textbf{\bibinfo{volume}{382}},
  \bibinfo{pages}{1416--1421}, \doiprefix\doilink{10.1126/science.adi2336}
  (\bibinfo{year}{2023}).
\newblock \bibinfo{note}{ArXiv:2212.12794}.

\bibitem{chen2023fengwu}
\bibinfo{author}{Chen, K.} \emph{et~al.}
\newblock \bibinfo{journal}{\bibinfo{title}{The operational medium-range
  deterministic weather forecasting can be extended beyond a 10-day lead
  time}}.
\newblock {\emph{\JournalTitle{Communications Earth \& Environment}}}
  \textbf{\bibinfo{volume}{6}}, \bibinfo{pages}{518},
  \doiprefix\doilink{10.1038/s43247-025-02502-y} (\bibinfo{year}{2025}).
\newblock \bibinfo{note}{FengWu; arXiv:2304.02948}.

\bibitem{hersbach2020era5}
\bibinfo{author}{Hersbach, H.} \emph{et~al.}
\newblock \bibinfo{journal}{\bibinfo{title}{The {ERA5} global reanalysis}}.
\newblock {\emph{\JournalTitle{Quarterly Journal of the Royal Meteorological
  Society}}} \textbf{\bibinfo{volume}{146}}, \bibinfo{pages}{1999--2049},
  \doiprefix\doilink{10.1002/qj.3803} (\bibinfo{year}{2020}).

\bibitem{carrassi2018da}
\bibinfo{author}{Carrassi, A.}, \bibinfo{author}{Bocquet, M.},
  \bibinfo{author}{Bertino, L.} \& \bibinfo{author}{Evensen, G.}
\newblock \bibinfo{journal}{\bibinfo{title}{Data assimilation in the
  geosciences: An overview of methods, issues, and perspectives}}.
\newblock {\emph{\JournalTitle{WIREs Climate Change}}}
  \textbf{\bibinfo{volume}{9}}, \bibinfo{pages}{e535},
  \doiprefix\doilink{10.1002/wcc.535} (\bibinfo{year}{2018}).

\bibitem{slivinski2025assimilating}
\bibinfo{author}{Slivinski, L.~C.}, \bibinfo{author}{Whitaker, J.~S.},
  \bibinfo{author}{Frolov, S.}, \bibinfo{author}{Smith, T.~A.} \&
  \bibinfo{author}{Agarwal, N.}
\newblock \bibinfo{journal}{\bibinfo{title}{Assimilating observed surface
  pressure into {ML} weather prediction models}}.
\newblock {\emph{\JournalTitle{Geophysical Research Letters}}}
  \textbf{\bibinfo{volume}{52}}, \bibinfo{pages}{e2024GL114396},
  \doiprefix\doilink{10.1029/2024GL114396} (\bibinfo{year}{2025}).

\bibitem{huang2026cycling}
\bibinfo{author}{Huang, H.}, \bibinfo{author}{Lei, L.} \& \bibinfo{author}{Tan,
  Z.-M.}
\newblock \bibinfo{journal}{\bibinfo{title}{Ensemble-based assimilation of
  sounding observations with {AI} weather models}}.
\newblock {\emph{\JournalTitle{Geophysical Research Letters}}}
  \textbf{\bibinfo{volume}{53}}, \bibinfo{pages}{e2026GL122153},
  \doiprefix\doilink{10.1029/2026GL122153} (\bibinfo{year}{2026}).

\bibitem{allen2025aardvark}
\bibinfo{author}{Allen, A.} \emph{et~al.}
\newblock \bibinfo{journal}{\bibinfo{title}{End-to-end data-driven weather
  prediction}}.
\newblock {\emph{\JournalTitle{Nature}}} \textbf{\bibinfo{volume}{641}},
  \bibinfo{pages}{1172--1179}, \doiprefix\doilink{10.1038/s41586-025-08897-0}
  (\bibinfo{year}{2025}).
\newblock \bibinfo{note}{ArXiv:2404.00411}.

\bibitem{alexe2024graphdop}
\bibinfo{author}{Alexe, M.} \emph{et~al.}
\newblock \bibinfo{journal}{\bibinfo{title}{{GraphDOP}: Towards skilful
  data-driven medium-range weather forecasts learnt and initialised directly
  from observations}}.
\newblock {\emph{\JournalTitle{arXiv preprint}}}
  \doiprefix\doilink{10.48550/arXiv.2412.15687} (\bibinfo{year}{2024}).

\bibitem{anna_vaughan_2025}
\bibinfo{author}{Vaughan, A.}
\newblock \bibinfo{title}{aardvark-weather (revision 1a29cb4)},
  \doiprefix\doilink{10.57967/hf/4274} (\bibinfo{year}{2025}).

\bibitem{murphy1977costloss}
\bibinfo{author}{Murphy, A.~H.}
\newblock \bibinfo{journal}{\bibinfo{title}{The value of climatological,
  categorical and probabilistic forecasts in the cost-loss ratio situation}}.
\newblock {\emph{\JournalTitle{Monthly Weather Review}}}
  \textbf{\bibinfo{volume}{105}}, \bibinfo{pages}{803--816},
  \doiprefix\doilink{10.1175/1520-0493(1977)105<0803:TVOCCA>2.0.CO;2}
  (\bibinfo{year}{1977}).

\bibitem{fundel2019dialogue}
\bibinfo{author}{Fundel, V.~J.}, \bibinfo{author}{Fleischhut, N.},
  \bibinfo{author}{Herzog, S.~M.}, \bibinfo{author}{G{\"o}ber, M.} \&
  \bibinfo{author}{Hagedorn, R.}
\newblock \bibinfo{journal}{\bibinfo{title}{Promoting the use of probabilistic
  weather forecasts through a dialogue between scientists, developers and
  end-users}}.
\newblock {\emph{\JournalTitle{Quarterly Journal of the Royal Meteorological
  Society}}} \textbf{\bibinfo{volume}{145}}, \bibinfo{pages}{210--231},
  \doiprefix\doilink{10.1002/qj.3482} (\bibinfo{year}{2019}).

\bibitem{benbouallegue2024rise}
\bibinfo{author}{Ben~Bouall{\`e}gue, Z.}, \bibinfo{author}{Clare, M. C.~A.},
  \bibinfo{author}{Magnusson, L.}, \bibinfo{author}{Gasc{\'o}n, E.}
  \emph{et~al.}
\newblock \bibinfo{journal}{\bibinfo{title}{The rise of data-driven weather
  forecasting: {A} first statistical assessment of machine learning-based
  weather forecasts in an operational-like context}}.
\newblock {\emph{\JournalTitle{Bulletin of the American Meteorological
  Society}}} \textbf{\bibinfo{volume}{105}}, \bibinfo{pages}{E864--E883},
  \doiprefix\doilink{10.1175/BAMS-D-23-0162.1} (\bibinfo{year}{2024}).

\bibitem{price2025gencast}
\bibinfo{author}{Price, I.} \emph{et~al.}
\newblock \bibinfo{journal}{\bibinfo{title}{Probabilistic weather forecasting
  with machine learning}}.
\newblock {\emph{\JournalTitle{Nature}}} \textbf{\bibinfo{volume}{637}},
  \bibinfo{pages}{84--90}, \doiprefix\doilink{10.1038/s41586-024-08252-9}
  (\bibinfo{year}{2025}).
\newblock \bibinfo{note}{GenCast; arXiv:2312.15796}.

\bibitem{abdar2021review}
\bibinfo{author}{Abdar, M.} \emph{et~al.}
\newblock \bibinfo{journal}{\bibinfo{title}{A review of uncertainty
  quantification in deep learning: Techniques, applications and challenges}}.
\newblock {\emph{\JournalTitle{Information Fusion}}}
  \textbf{\bibinfo{volume}{76}}, \bibinfo{pages}{243--297},
  \doiprefix\doilink{10.1016/j.inffus.2021.05.008} (\bibinfo{year}{2021}).
\newblock \bibinfo{note}{ArXiv:2011.06225}.

\bibitem{almeidaPredictiveSkillArtificial2026}
\bibinfo{author}{Almeida, R.}, \bibinfo{author}{Otero, N.},
  \bibinfo{author}{{Fern\'andez-Torres}, M.-{\'A}.} \& \bibinfo{author}{Ma, J.}
\newblock \bibinfo{journal}{\bibinfo{title}{On the {{Predictive Skill}} of
  {{Artificial Intelligence-based Weather Models}} for {{Extreme Events}} using
  {{Uncertainty Quantification}}}}.
\newblock {\emph{\JournalTitle{Artificial Intelligence for the Earth Systems}}}
  \textbf{\bibinfo{volume}{5}}, \doiprefix\doilink{10.1175/AIES-D-25-0113.1}
  (\bibinfo{year}{2026}).

\bibitem{buelte2024uq}
\bibinfo{author}{B{\"u}lte, C.}, \bibinfo{author}{Horat, N.},
  \bibinfo{author}{Quinting, J.} \& \bibinfo{author}{Lerch, S.}
\newblock \bibinfo{journal}{\bibinfo{title}{Uncertainty quantification for
  data-driven weather models}}.
\newblock {\emph{\JournalTitle{Artificial Intelligence for the Earth Systems}}}
  \textbf{\bibinfo{volume}{5}}, \doiprefix\doilink{10.1175/AIES-D-24-0049.1}
  (\bibinfo{year}{2026}).
\newblock \bibinfo{note}{ArXiv:2403.13458}.

\bibitem{langAIFSCRPSEnsembleForecasting2026}
\bibinfo{author}{Lang, S.} \emph{et~al.}
\newblock \bibinfo{journal}{\bibinfo{title}{{{AIFS-CRPS}}: Ensemble forecasting
  using a model trained with a loss function based on the continuous ranked
  probability score}}.
\newblock {\emph{\JournalTitle{npj Artificial Intelligence}}}
  \textbf{\bibinfo{volume}{2}}, \bibinfo{pages}{18},
  \doiprefix\doilink{10.1038/s44387-026-00073-7} (\bibinfo{year}{2026}).

\bibitem{leutbecher2008ensemble}
\bibinfo{author}{Leutbecher, M.} \& \bibinfo{author}{Palmer, T.~N.}
\newblock \bibinfo{journal}{\bibinfo{title}{Ensemble forecasting}}.
\newblock {\emph{\JournalTitle{Journal of Computational Physics}}}
  \textbf{\bibinfo{volume}{227}}, \bibinfo{pages}{3515--3539},
  \doiprefix\doilink{10.1016/j.jcp.2007.02.014} (\bibinfo{year}{2008}).

\bibitem{molteni1996eps}
\bibinfo{author}{Molteni, F.}, \bibinfo{author}{Buizza, R.},
  \bibinfo{author}{Palmer, T.~N.} \& \bibinfo{author}{Petroliagis, T.}
\newblock \bibinfo{journal}{\bibinfo{title}{The {ECMWF} ensemble prediction
  system: Methodology and validation}}.
\newblock {\emph{\JournalTitle{Quarterly Journal of the Royal Meteorological
  Society}}} \textbf{\bibinfo{volume}{122}}, \bibinfo{pages}{73--119},
  \doiprefix\doilink{10.1002/qj.49712252905} (\bibinfo{year}{1996}).

\bibitem{buizza1995singular}
\bibinfo{author}{Buizza, R.} \& \bibinfo{author}{Palmer, T.~N.}
\newblock \bibinfo{journal}{\bibinfo{title}{The singular-vector structure of
  the atmospheric global circulation}}.
\newblock {\emph{\JournalTitle{Journal of the Atmospheric Sciences}}}
  \textbf{\bibinfo{volume}{52}}, \bibinfo{pages}{1434--1456},
  \doiprefix\doilink{10.1175/1520-0469(1995)052<1434:TSVSOT>2.0.CO;2}
  (\bibinfo{year}{1995}).

\bibitem{toth1997breeding}
\bibinfo{author}{Toth, Z.} \& \bibinfo{author}{Kalnay, E.}
\newblock \bibinfo{journal}{\bibinfo{title}{Ensemble forecasting at {NCEP} and
  the breeding method}}.
\newblock {\emph{\JournalTitle{Monthly Weather Review}}}
  \textbf{\bibinfo{volume}{125}}, \bibinfo{pages}{3297--3319},
  \doiprefix\doilink{10.1175/1520-0493(1997)125<3297:EFANAT>2.0.CO;2}
  (\bibinfo{year}{1997}).

\bibitem{isaksen2010eda}
\bibinfo{author}{Isaksen, L.} \emph{et~al.}
\newblock \bibinfo{title}{Ensemble of data assimilations at {ECMWF}}.
\newblock \bibinfo{type}{Technical Memorandum} \bibinfo{number}{636},
  \bibinfo{institution}{ECMWF} (\bibinfo{year}{2010}).
\newblock \doiprefix\doilink{10.21957/obke4k60}.

\bibitem{bonavita2012eda}
\bibinfo{author}{Bonavita, M.}, \bibinfo{author}{Isaksen, L.} \&
  \bibinfo{author}{H{\'o}lm, E.}
\newblock \bibinfo{journal}{\bibinfo{title}{On the use of {EDA} background
  error variances in the {ECMWF 4D-Var}}}.
\newblock {\emph{\JournalTitle{Quarterly Journal of the Royal Meteorological
  Society}}} \textbf{\bibinfo{volume}{138}}, \bibinfo{pages}{1540--1559},
  \doiprefix\doilink{10.1002/qj.1899} (\bibinfo{year}{2012}).

\bibitem{palmer2009stochastic}
\bibinfo{author}{Palmer, T.~N.} \emph{et~al.}
\newblock \bibinfo{title}{Stochastic parametrization and model uncertainty}.
\newblock \bibinfo{type}{Technical Memorandum} \bibinfo{number}{598},
  \bibinfo{institution}{ECMWF} (\bibinfo{year}{2009}).
\newblock \doiprefix\doilink{10.21957/ps8gbwbdv}.

\bibitem{leutbecher2017stochastic}
\bibinfo{author}{Leutbecher, M.} \emph{et~al.}
\newblock \bibinfo{journal}{\bibinfo{title}{Stochastic representations of model
  uncertainties at {ECMWF}: state of the art and future vision}}.
\newblock {\emph{\JournalTitle{Quarterly Journal of the Royal Meteorological
  Society}}} \textbf{\bibinfo{volume}{143}}, \bibinfo{pages}{2315--2339},
  \doiprefix\doilink{10.1002/qj.3094} (\bibinfo{year}{2017}).

\bibitem{derkiureghian2009aleatory}
\bibinfo{author}{Der~Kiureghian, A.} \& \bibinfo{author}{Ditlevsen, O.}
\newblock \bibinfo{journal}{\bibinfo{title}{Aleatory or epistemic? {Does} it
  matter?}}
\newblock {\emph{\JournalTitle{Structural Safety}}}
  \textbf{\bibinfo{volume}{31}}, \bibinfo{pages}{105--112},
  \doiprefix\doilink{10.1016/j.strusafe.2008.06.020} (\bibinfo{year}{2009}).

\bibitem{kendall2017uncertainties}
\bibinfo{author}{Kendall, A.} \& \bibinfo{author}{Gal, Y.}
\newblock \bibinfo{title}{What uncertainties do we need in {Bayesian} deep
  learning for computer vision?}
\newblock In \emph{\bibinfo{booktitle}{Advances in Neural Information
  Processing Systems (NeurIPS)}},
  \urlprefix\url{https://proceedings.neurips.cc/paper/2017/hash/2650d6089a6d640c5e85b2b88265dc2b-Abstract.html}
  (\bibinfo{year}{2017}).

\bibitem{huellermeier2021aleatoric}
\bibinfo{author}{H{\"u}llermeier, E.} \& \bibinfo{author}{Waegeman, W.}
\newblock \bibinfo{journal}{\bibinfo{title}{Aleatoric and epistemic uncertainty
  in machine learning: an introduction to concepts and methods}}.
\newblock {\emph{\JournalTitle{Machine Learning}}}
  \textbf{\bibinfo{volume}{110}}, \bibinfo{pages}{457--506},
  \doiprefix\doilink{10.1007/s10994-021-05946-3} (\bibinfo{year}{2021}).

\bibitem{baur2026disentanglement}
\bibinfo{author}{Baur, S.}, \bibinfo{author}{Samek, W.} \& \bibinfo{author}{Ma,
  J.}
\newblock \bibinfo{title}{Benchmarking uncertainty and its disentanglement in
  multi-label chest {X}-ray classification}.
\newblock In \emph{\bibinfo{booktitle}{Uncertainty for Safe Utilization of
  Machine Learning in Medical Imaging (UNSURE)}}, vol. \bibinfo{volume}{16166}
  of \emph{\bibinfo{series}{Lecture Notes in Computer Science}},
  \bibinfo{pages}{193--203}, \doiprefix\doilink{10.1007/978-3-032-06593-3_18}
  (\bibinfo{publisher}{Springer}, \bibinfo{year}{2025}).

\bibitem{baur2026beyond}
\bibinfo{author}{Baur, S.}, \bibinfo{author}{Schernich, A.},
  \bibinfo{author}{B{\"o}ke, E.}, \bibinfo{author}{Samek, W.} \&
  \bibinfo{author}{Ma, J.}
\newblock \bibinfo{journal}{\bibinfo{title}{Beyond boundary noise: Aggregated
  aleatoric uncertainty fails to capture presence ambiguity in {3D} lung nodule
  segmentation}}.
\newblock {\emph{\JournalTitle{arXiv preprint arXiv:2608.14766}}}
  (\bibinfo{year}{2026}).

\bibitem{depeweg2018decomposition}
\bibinfo{author}{Depeweg, S.}, \bibinfo{author}{Hern{\'a}ndez-Lobato, J.~M.},
  \bibinfo{author}{Doshi-Velez, F.} \& \bibinfo{author}{Udluft, S.}
\newblock \bibinfo{title}{Decomposition of uncertainty in {Bayesian} deep
  learning for efficient and risk-sensitive learning}.
\newblock In \emph{\bibinfo{booktitle}{Proceedings of the 35th International
  Conference on Machine Learning (ICML)}}, vol.~\bibinfo{volume}{80} of
  \emph{\bibinfo{series}{PMLR}},
  \urlprefix\url{https://proceedings.mlr.press/v80/depeweg18a.html}
  (\bibinfo{year}{2018}).

\bibitem{valdenegro2022deeper}
\bibinfo{author}{Valdenegro-Toro, M.} \& \bibinfo{author}{Mori, D.~S.}
\newblock \bibinfo{title}{A deeper look into aleatoric and epistemic
  uncertainty disentanglement}.
\newblock In \emph{\bibinfo{booktitle}{IEEE/CVF Conference on Computer Vision
  and Pattern Recognition Workshops (CVPRW)}}, \bibinfo{pages}{1508--1516},
  \doiprefix\doilink{10.1109/CVPRW56347.2022.00157} (\bibinfo{year}{2022}).

\bibitem{houlsby2011bald}
\bibinfo{author}{Houlsby, N.}, \bibinfo{author}{Husz{\'a}r, F.},
  \bibinfo{author}{Ghahramani, Z.} \& \bibinfo{author}{Lengyel, M.}
\newblock \bibinfo{journal}{\bibinfo{title}{Bayesian active learning for
  classification and preference learning}}.
\newblock {\emph{\JournalTitle{arXiv preprint}}}
  \doiprefix\doilink{10.48550/arXiv.1112.5745} (\bibinfo{year}{2011}).

\bibitem{smith2018understanding}
\bibinfo{author}{Smith, L.} \& \bibinfo{author}{Gal, Y.}
\newblock \bibinfo{title}{Understanding measures of uncertainty for adversarial
  example detection}.
\newblock In \emph{\bibinfo{booktitle}{Proceedings of the 34th Conference on
  Uncertainty in Artificial Intelligence (UAI)}},
  \urlprefix\url{http://auai.org/uai2018/proceedings/papers/207.pdf}
  (\bibinfo{year}{2018}).

\bibitem{mansfieldEpistemicAleatoricUncertainty}
\bibinfo{author}{Mansfield, L.~A.} \& \bibinfo{author}{Christensen, H.~M.}
\newblock \bibinfo{journal}{\bibinfo{title}{Epistemic and aleatoric uncertainty
  quantification in weather and climate models}}.
\newblock {\emph{\JournalTitle{Quarterly Journal of the Royal Meteorological
  Society}}} \textbf{\bibinfo{volume}{152}}, \bibinfo{pages}{e70219},
  \doiprefix\doilink{10.1002/qj.70219} (\bibinfo{year}{2026}).

\bibitem{haynesCreatingEvaluatingUncertainty2023}
\bibinfo{author}{Haynes, K.}, \bibinfo{author}{Lagerquist, R.},
  \bibinfo{author}{McGraw, M.}, \bibinfo{author}{Musgrave, K.} \&
  \bibinfo{author}{{Ebert-Uphoff}, I.}
\newblock \bibinfo{journal}{\bibinfo{title}{Creating and {{Evaluating
  Uncertainty Estimates}} with {{Neural Networks}} for {{Environmental-Science
  Applications}}}}.
\newblock {\emph{\JournalTitle{Artificial Intelligence for the Earth Systems}}}
  \textbf{\bibinfo{volume}{2}}, \doiprefix\doilink{10.1175/AIES-D-22-0061.1}
  (\bibinfo{year}{2023}).

\bibitem{hersbach2000crps}
\bibinfo{author}{Hersbach, H.}
\newblock \bibinfo{journal}{\bibinfo{title}{Decomposition of the continuous
  ranked probability score for ensemble prediction systems}}.
\newblock {\emph{\JournalTitle{Weather and Forecasting}}}
  \textbf{\bibinfo{volume}{15}}, \bibinfo{pages}{559--570},
  \doiprefix\doilink{10.1175/1520-0434(2000)015<0559:DOTCRP>2.0.CO;2}
  (\bibinfo{year}{2000}).

\bibitem{gneiting2007scoring}
\bibinfo{author}{Gneiting, T.} \& \bibinfo{author}{Raftery, A.~E.}
\newblock \bibinfo{journal}{\bibinfo{title}{Strictly proper scoring rules,
  prediction, and estimation}}.
\newblock {\emph{\JournalTitle{Journal of the American Statistical
  Association}}} \textbf{\bibinfo{volume}{102}}, \bibinfo{pages}{359--378},
  \doiprefix\doilink{10.1198/016214506000001437} (\bibinfo{year}{2007}).

\bibitem{gneiting2007calibration}
\bibinfo{author}{Gneiting, T.}, \bibinfo{author}{Balabdaoui, F.} \&
  \bibinfo{author}{Raftery, A.~E.}
\newblock \bibinfo{journal}{\bibinfo{title}{Probabilistic forecasts,
  calibration and sharpness}}.
\newblock {\emph{\JournalTitle{Journal of the Royal Statistical Society: Series
  B}}} \textbf{\bibinfo{volume}{69}}, \bibinfo{pages}{243--268},
  \doiprefix\doilink{10.1111/j.1467-9868.2007.00587.x} (\bibinfo{year}{2007}).

\bibitem{hamill2001rank}
\bibinfo{author}{Hamill, T.~M.}
\newblock \bibinfo{journal}{\bibinfo{title}{Interpretation of rank histograms
  for verifying ensemble forecasts}}.
\newblock {\emph{\JournalTitle{Monthly Weather Review}}}
  \textbf{\bibinfo{volume}{129}}, \bibinfo{pages}{550--560},
  \doiprefix\doilink{10.1175/1520-0493(2001)129<0550:IORHFV>2.0.CO;2}
  (\bibinfo{year}{2001}).

\bibitem{gal2016dropout}
\bibinfo{author}{Gal, Y.} \& \bibinfo{author}{Ghahramani, Z.}
\newblock \bibinfo{title}{Dropout as a {Bayesian} approximation: Representing
  model uncertainty in deep learning}.
\newblock In \emph{\bibinfo{booktitle}{Proceedings of the 33rd International
  Conference on Machine Learning (ICML)}},
  \urlprefix\url{https://proceedings.mlr.press/v48/gal16.html}
  (\bibinfo{year}{2016}).

\bibitem{cachay2026ucast}
\bibinfo{author}{R{\"u}hling~Cachay, S.}, \bibinfo{author}{Watson-Parris, D.}
  \& \bibinfo{author}{Yu, R.}
\newblock \bibinfo{title}{{U-Cast}: A surprisingly simple and efficient
  frontier probabilistic {AI} weather forecaster}.
\newblock In \emph{\bibinfo{booktitle}{Forty-third International Conference on
  Machine Learning (ICML)}},
  \urlprefix\url{https://openreview.net/forum?id=XF0wkyEbuM}
  (\bibinfo{year}{2026}).

\bibitem{bley2025secondorder}
\bibinfo{author}{Bley, F.}, \bibinfo{author}{Lapuschkin, S.},
  \bibinfo{author}{Samek, W.} \& \bibinfo{author}{Montavon, G.}
\newblock \bibinfo{journal}{\bibinfo{title}{Explaining predictive uncertainty
  by exposing second-order effects}}.
\newblock {\emph{\JournalTitle{Pattern Recognition}}}
  \textbf{\bibinfo{volume}{160}}, \bibinfo{pages}{111171},
  \doiprefix\doilink{10.1016/j.patcog.2024.111171} (\bibinfo{year}{2025}).

\bibitem{dunn2012hadisd}
\bibinfo{author}{Dunn, R. J.~H.} \emph{et~al.}
\newblock \bibinfo{journal}{\bibinfo{title}{{HadISD}: a quality-controlled
  global synoptic report database for selected variables at long-term stations
  from 1973--2011}}.
\newblock {\emph{\JournalTitle{Climate of the Past}}}
  \textbf{\bibinfo{volume}{8}}, \bibinfo{pages}{1649--1679},
  \doiprefix\doilink{10.5194/cp-8-1649-2012} (\bibinfo{year}{2012}).

\bibitem{dramsch2025explainability}
\bibinfo{author}{Dramsch, J.~S.}, \bibinfo{author}{Kuglitsch, M.~M.},
  \bibinfo{author}{Fern{\'a}ndez-Torres, M.-{\'A}.} \emph{et~al.}
\newblock \bibinfo{journal}{\bibinfo{title}{Explainability can foster trust in
  artificial intelligence in geoscience}}.
\newblock {\emph{\JournalTitle{Nature Geoscience}}}
  \textbf{\bibinfo{volume}{18}}, \bibinfo{pages}{112--114},
  \doiprefix\doilink{10.1038/s41561-025-01639-x} (\bibinfo{year}{2025}).

\bibitem{bormann2019ose}
\bibinfo{author}{Bormann, N.}, \bibinfo{author}{Lawrence, H.} \&
  \bibinfo{author}{Farnan, J.}
\newblock \bibinfo{title}{Global observing system experiments in the {ECMWF}
  assimilation system}.
\newblock \bibinfo{type}{Technical Memorandum} \bibinfo{number}{839},
  \bibinfo{institution}{ECMWF} (\bibinfo{year}{2019}).
\newblock \doiprefix\doilink{10.21957/sr184iyz}.

\bibitem{sandu2013stable}
\bibinfo{author}{Sandu, I.}, \bibinfo{author}{Beljaars, A.},
  \bibinfo{author}{Bechtold, P.}, \bibinfo{author}{Mauritsen, T.} \&
  \bibinfo{author}{Balsamo, G.}
\newblock \bibinfo{journal}{\bibinfo{title}{Why is it so difficult to represent
  stably stratified conditions in numerical weather prediction ({NWP})
  models?}}
\newblock {\emph{\JournalTitle{Journal of Advances in Modeling Earth Systems}}}
  \textbf{\bibinfo{volume}{5}}, \bibinfo{pages}{117--133},
  \doiprefix\doilink{10.1002/jame.20013} (\bibinfo{year}{2013}).

\bibitem{wallace1989sstwind}
\bibinfo{author}{Wallace, J.~M.}, \bibinfo{author}{Mitchell, T.~P.} \&
  \bibinfo{author}{Deser, C.}
\newblock \bibinfo{journal}{\bibinfo{title}{The influence of sea-surface
  temperature on surface wind in the eastern equatorial {Pacific}: Seasonal and
  interannual variability}}.
\newblock {\emph{\JournalTitle{Journal of Climate}}}
  \textbf{\bibinfo{volume}{2}}, \bibinfo{pages}{1492--1499},
  \doiprefix\doilink{10.1175/1520-0442(1989)002<1492:TIOSST>2.0.CO;2}
  (\bibinfo{year}{1989}).

\bibitem{barsugli1998coupling}
\bibinfo{author}{Barsugli, J.~J.} \& \bibinfo{author}{Battisti, D.~S.}
\newblock \bibinfo{journal}{\bibinfo{title}{The basic effects of
  atmosphere--ocean thermal coupling on midlatitude variability}}.
\newblock {\emph{\JournalTitle{Journal of the Atmospheric Sciences}}}
  \textbf{\bibinfo{volume}{55}}, \bibinfo{pages}{477--493},
  \doiprefix\doilink{10.1175/1520-0469(1998)055<0477:TBEOAO>2.0.CO;2}
  (\bibinfo{year}{1998}).

\bibitem{zagar2017limits}
\bibinfo{author}{{\v{Z}}agar, N.}
\newblock \bibinfo{journal}{\bibinfo{title}{A global perspective of the limits
  of prediction skill of {NWP} models}}.
\newblock {\emph{\JournalTitle{Tellus A: Dynamic Meteorology and
  Oceanography}}} \textbf{\bibinfo{volume}{69}}, \bibinfo{pages}{1317573},
  \doiprefix\doilink{10.1080/16000870.2017.1317573} (\bibinfo{year}{2017}).

\bibitem{bouttier2001ose}
\bibinfo{author}{Bouttier, F.} \& \bibinfo{author}{Kelly, G.}
\newblock \bibinfo{journal}{\bibinfo{title}{Observing-system experiments in the
  {ECMWF} {4D-Var} data assimilation system}}.
\newblock {\emph{\JournalTitle{Quarterly Journal of the Royal Meteorological
  Society}}} \textbf{\bibinfo{volume}{127}}, \bibinfo{pages}{1469--1488},
  \doiprefix\doilink{10.1002/qj.49712757419} (\bibinfo{year}{2001}).

\bibitem{langland2004fsoi}
\bibinfo{author}{Langland, R.~H.} \& \bibinfo{author}{Baker, N.~L.}
\newblock \bibinfo{journal}{\bibinfo{title}{Estimation of observation impact
  using the {NRL} atmospheric variational data assimilation adjoint system}}.
\newblock {\emph{\JournalTitle{Tellus A}}} \textbf{\bibinfo{volume}{56}},
  \bibinfo{pages}{189--201},
  \doiprefix\doilink{10.1111/j.1600-0870.2004.00056.x} (\bibinfo{year}{2004}).

\bibitem{gal2017concrete}
\bibinfo{author}{Gal, Y.}, \bibinfo{author}{Hron, J.} \&
  \bibinfo{author}{Kendall, A.}
\newblock \bibinfo{title}{Concrete dropout}.
\newblock In \emph{\bibinfo{booktitle}{Advances in Neural Information
  Processing Systems (NeurIPS)}},
  \urlprefix\url{https://proceedings.neurips.cc/paper/2017/hash/84ddfb34126fc3a48ee38d7044e87276-Abstract.html}
  (\bibinfo{year}{2017}).

\bibitem{lakshminarayanan2017deepensembles}
\bibinfo{author}{Lakshminarayanan, B.}, \bibinfo{author}{Pritzel, A.} \&
  \bibinfo{author}{Blundell, C.}
\newblock \bibinfo{title}{Simple and scalable predictive uncertainty estimation
  using deep ensembles}.
\newblock In \emph{\bibinfo{booktitle}{Advances in Neural Information
  Processing Systems (NeurIPS)}},
  \urlprefix\url{https://proceedings.neurips.cc/paper/2017/hash/9ef2ed4b7fd2c810847ffa5fa85bce38-Abstract.html}
  (\bibinfo{year}{2017}).

\bibitem{dosovitskiy2021vit}
\bibinfo{author}{Dosovitskiy, A.} \emph{et~al.}
\newblock \bibinfo{title}{An image is worth 16x16 words: Transformers for image
  recognition at scale}.
\newblock In \emph{\bibinfo{booktitle}{International Conference on Learning
  Representations (ICLR)}},
  \urlprefix\url{https://openreview.net/forum?id=YicbFdNTTy}
  (\bibinfo{year}{2021}).

\bibitem{ronneberger2015unet}
\bibinfo{author}{Ronneberger, O.}, \bibinfo{author}{Fischer, P.} \&
  \bibinfo{author}{Brox, T.}
\newblock \bibinfo{title}{{U-Net}: Convolutional networks for biomedical image
  segmentation}.
\newblock In \emph{\bibinfo{booktitle}{Medical Image Computing and
  Computer-Assisted Intervention (MICCAI)}}, \bibinfo{pages}{234--241},
  \doiprefix\doilink{10.1007/978-3-319-24574-4_28} (\bibinfo{year}{2015}).

\bibitem{perez2018film}
\bibinfo{author}{Perez, E.}, \bibinfo{author}{Strub, F.}, \bibinfo{author}{{de
  Vries}, H.}, \bibinfo{author}{Dumoulin, V.} \& \bibinfo{author}{Courville,
  A.}
\newblock \bibinfo{title}{{FiLM}: Visual reasoning with a general conditioning
  layer}.
\newblock In \emph{\bibinfo{booktitle}{Proceedings of the AAAI Conference on
  Artificial Intelligence}}, vol.~\bibinfo{volume}{32},
  \doiprefix\doilink{10.1609/aaai.v32i1.11671} (\bibinfo{year}{2018}).

\bibitem{wilks2018enforcing}
\bibinfo{author}{Wilks, D.~S.}
\newblock \bibinfo{journal}{\bibinfo{title}{Enforcing calibration in ensemble
  postprocessing}}.
\newblock {\emph{\JournalTitle{Quarterly Journal of the Royal Meteorological
  Society}}} \textbf{\bibinfo{volume}{144}}, \bibinfo{pages}{76--84},
  \doiprefix\doilink{10.1002/qj.3185} (\bibinfo{year}{2018}).

\bibitem{rasp2024weatherbench2}
\bibinfo{author}{Rasp, S.} \emph{et~al.}
\newblock \bibinfo{journal}{\bibinfo{title}{{WeatherBench 2}: A benchmark for
  the next generation of data-driven global weather models}}.
\newblock {\emph{\JournalTitle{Journal of Advances in Modeling Earth Systems}}}
  \textbf{\bibinfo{volume}{16}}, \bibinfo{pages}{e2023MS004019},
  \doiprefix\doilink{10.1029/2023MS004019} (\bibinfo{year}{2024}).

\bibitem{weatherbenchx2024}
\bibinfo{author}{{Google Research}}.
\newblock \bibinfo{title}{{WeatherBench-X}: a modular framework for evaluating
  weather forecasts}.
\newblock \bibinfo{howpublished}{GitHub repository,
  \url{https://github.com/google-research/weatherbenchX}}
  (\bibinfo{year}{2024}).
\newblock \bibinfo{note}{Apache-2.0 licence; accessed November 2026}.

\bibitem{richardson2000value}
\bibinfo{author}{Richardson, D.~S.}
\newblock \bibinfo{journal}{\bibinfo{title}{Skill and relative economic value
  of the {ECMWF} ensemble prediction system}}.
\newblock {\emph{\JournalTitle{Quarterly Journal of the Royal Meteorological
  Society}}} \textbf{\bibinfo{volume}{126}}, \bibinfo{pages}{649--667},
  \doiprefix\doilink{10.1002/qj.49712656313} (\bibinfo{year}{2000}).

\bibitem{diebold1995comparing}
\bibinfo{author}{Diebold, F.~X.} \& \bibinfo{author}{Mariano, R.~S.}
\newblock \bibinfo{journal}{\bibinfo{title}{Comparing predictive accuracy}}.
\newblock {\emph{\JournalTitle{Journal of Business \& Economic Statistics}}}
  \textbf{\bibinfo{volume}{13}}, \bibinfo{pages}{253--263},
  \doiprefix\doilink{10.1080/07350015.1995.10524599} (\bibinfo{year}{1995}).

\bibitem{geer2016significance}
\bibinfo{author}{Geer, A.~J.}
\newblock \bibinfo{journal}{\bibinfo{title}{Significance of changes in
  medium-range forecast scores}}.
\newblock {\emph{\JournalTitle{Tellus A: Dynamic Meteorology and
  Oceanography}}} \textbf{\bibinfo{volume}{68}}, \bibinfo{pages}{30229},
  \doiprefix\doilink{10.3402/tellusa.v68.30229} (\bibinfo{year}{2016}).

\end{thebibliography}

\section*{Acknowledgements}

This research has been supported by the European Union's Horizon Europe research
and innovation program (EU Horizon Europe) project MedEWSa under grant agreement
no. 101121192 and ARTEMis under grant agreement no. 101225852.

\section*{Author contributions statement}

R.A.\ conceived the study, developed the uncertainty disentanglement framework,
implemented the probabilistic models, performed the training, experiments and
analysis, and wrote the manuscript. N.O.\ contributed to the conception and
design of the study. J.A.\ contributed to the model training and the development
of the framework. S.B.\ contributed to the uncertainty quantification
fundamentals and the disentanglement methodology. W.S.\ and J.M.\ supervised the
work. All authors discussed the results and reviewed the manuscript.

\section*{Additional information}

\textbf{Competing interests}: The authors declare no competing interests.

\clearpage
\renewcommand{\thefigure}{S\arabic{figure}}
\setcounter{figure}{0}
\renewcommand{\thetable}{S\arabic{table}}
\setcounter{table}{0}
\newcounter{suppnote}
\newcommand{\suppnote}[2]{\ifnum\value{suppnote}>0\clearpage\fi\refstepcounter{suppnote}%
	\subsection*{Supplementary Note \arabic{suppnote}: #1}\label{#2}}

\section*{Supplementary information}

\suppnote{Norm conditioning vs.\ embedding injection}{note:variants}

The injection sites of the two encoder noise-injection variants are sketched in
Supplementary Figure~\ref{fig:supp-variants}a; Supplementary
Figure~\ref{fig:supp-variants}b--d compares the two full chains (each a
noise-conditioned encoder plus a processor finetuned on that encoder's initial
conditions) under the evaluation protocol of the main text. The $4$--$5\,\%$
CRPS advantage of norm conditioning at days~1--4 decays monotonically and closes
to parity by day~10 (Supplementary Fig.~\ref{fig:supp-variants}b), with
near-identical total calibration (day-1 SSR $1.21$ vs.\ $1.27$) and attribution
($\rho_{\mathrm{drop}}=0.96$ at day~10; Supplementary
Fig.~\ref{fig:supp-variants}c,d); the difference is concentrated in the encoder
branch, where embedding injection produces a visibly larger encoder spread
(day-1 encoder-branch SSR $0.56$ vs.\ $0.46$; encoder mutual information $0.11$
vs.\ $0.09$\,nats) yet the worse mean state. The same pattern holds already at
lead~0, where the norm-conditioned encoder yields the better T2M analysis (RMSE
$1.02$ vs.\ $1.06$\,K, CRPS $0.54$ vs.\ $0.56$\,K) with \emph{less} spread
($0.82$ vs.\ $0.93$\,K). The two variants spend their perturbation budget
differently: embedding injection's single additive perturbation must be large
enough to survive all $L$ pretrained transformer blocks, whose LayerNorms and
attention were trained on unperturbed statistics, and this one-shot displacement
moves the encoder off its pretrained manifold, degrading the mean state; norm
conditioning composes the same calibrated output spread from many small,
per-feature and per-depth FiLM modulations that keep activations close to their
pretrained statistics, and its zero-initialized projections let finetuning
depart from the pretrained optimum gradually rather than from an
already-perturbed state. Consistent with this reading, the norm variant needs
less raw encoder spread for near-identical total calibration, and its skill
advantage is largest where the encoder perturbation dominates the forecast
(short leads), vanishing as processor uncertainty takes over (day~10,
Supplementary Fig.~\ref{fig:supp-variants}d).

\setlength{\textfloatsep}{6pt plus 2pt minus 4pt}
\begin{figure}[!htb]
	\centering
	\includegraphics{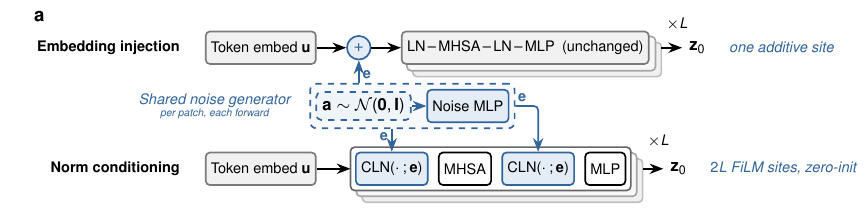}
	\par\vspace{1pt}
	\includegraphics[width=0.94\linewidth]{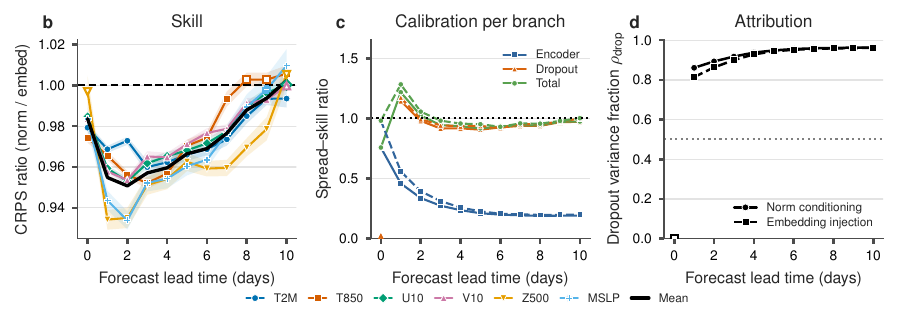}
	\caption{The two encoder noise-injection variants and their comparison.
		\textbf{a}, Both variants share one noise generator: per-patch Gaussian
		noise $\mathbf{a}$, redrawn at every forward pass, is mapped by a
		learned MLP to a per-patch embedding $\mathbf{e}$, the encoder-noise
		variable of Figure~\ref{fig:method}a. Embedding injection adds
		$\mathbf{e}$ once before the $L$ unchanged pretrained transformer
		blocks; norm conditioning re-injects the same realization at every
		block through zero-initialized conditional LayerNorms
		(CLN)\cite{perez2018film}.
		\textbf{b}--\textbf{d}, The two variants at equal ensemble size on
		the grid; lead 0 is the encoder analysis, so the dropout branch in
		\textbf{c} and $\rho_{\mathrm{drop}}$ in \textbf{d} are zero by
		construction. \textbf{b}, Fair-CRPS ratio per headline variable and
		their mean (black); values below the dashed parity line favor norm
		conditioning. \textbf{c}, Spread--skill ratio per branch, averaged
		over the headline variables, for norm conditioning (solid) and
		embedding injection (dashed). \textbf{d}, Dropout variance fraction
		$\rho_{\mathrm{drop}}$ (mean over variables) for both variants; the
		dotted lines mark perfect calibration (\textbf{c}) and parity between
		the branches (\textbf{d}).}
	\label{fig:supp-variants}
\end{figure}

\suppnote{Spread--skill and calibration}{note:calibration}

\begin{figure}[!htb]
	\centering
	\includegraphics[width=\linewidth]{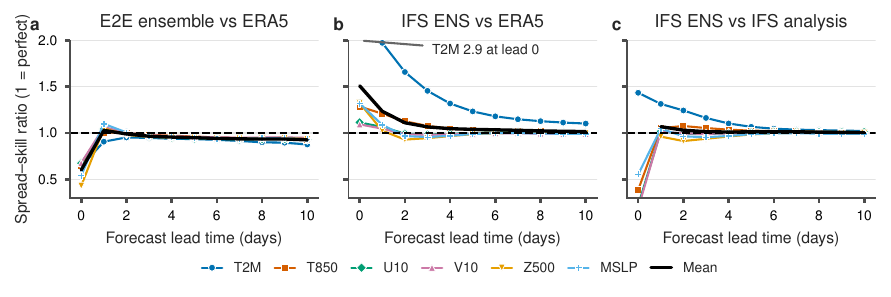}
	\caption{Spread--skill ratio per headline variable on the grid.
		\textbf{a}, The E2E ensemble from the analysis state (lead time 0) to
		day~10. \textbf{b}, IFS ENS from the WeatherBench~2 scores on the same
		axes; the published scores carry no per-initialization information, so
		no interval is shown. \textbf{c}, As \textbf{b}, but verified against
		the IFS operational analysis instead of ERA5.}
	\label{fig:supp-ssr}
\end{figure}

\begin{figure}[!htb]
	\centering
	\includegraphics[width=\linewidth]{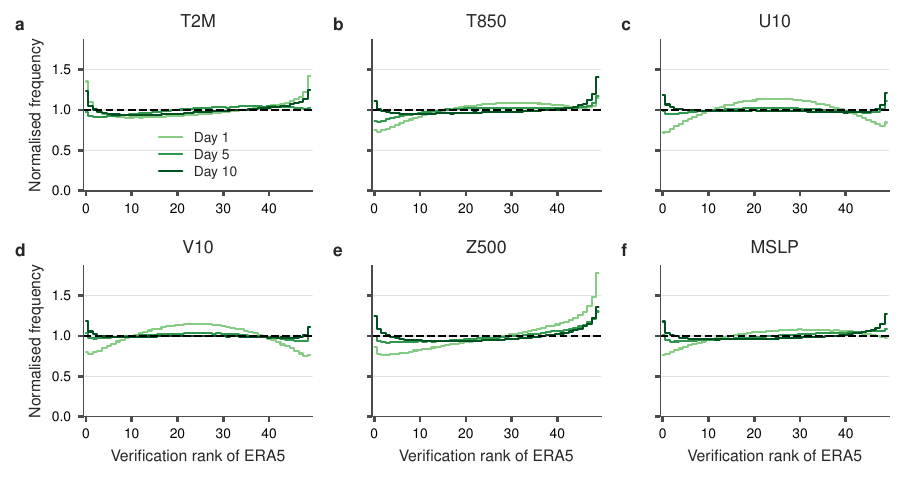}
	\caption{Gridded verification-rank histograms of the E2E ensemble
		for the six headline variables (\textbf{a}--\textbf{f}) at forecast
		days 1, 5 and 10; the dashed line marks a calibrated ensemble.}
	\label{fig:supp-rank-hist-gridded}
\end{figure}

\begin{figure}[!htb]
	\centering
	\includegraphics[width=0.68\linewidth]{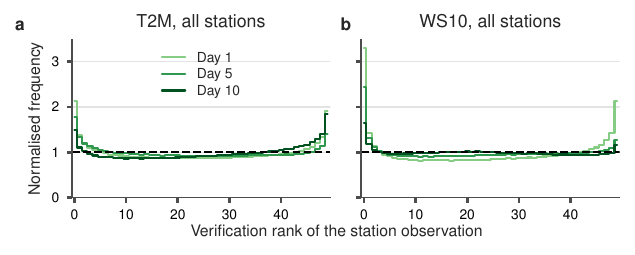}
	\caption{Station verification-rank histograms of the E2E ensemble of
		Figure~\ref{fig:stations} for T2M (\textbf{a}) and WS10 (\textbf{b})
		at forecast days 1, 5 and 10, over all verified HadISD stations; the
		dashed line marks a calibrated ensemble.}
	\label{fig:supp-rank-hist}
\end{figure}

Supplementary Figure~\ref{fig:supp-ssr} resolves the variable-mean
spread--skill ratio of Figure~\ref{fig:skill}d by headline variable. All six
variables are under-dispersive at analysis time ($0.43$ for Z500 to $0.68$ for
V10), close to calibrated at day~1 ($0.91$ for T2M to $1.10$ for MSLP) and
drift slowly below one, with day~1 to 10 means of $0.96$ to $0.97$ for every
variable but T2M ($0.92$). IFS ENS is within a few
percent of one from day~3 for every variable except T2M ($1.97$ at day~1
against ERA5), and differs most from its own-analysis ratios at short leads.
The verification-rank histograms complete the calibration check. On the grid
(Supplementary Fig.~\ref{fig:supp-rank-hist-gridded}) they are close to flat at
every lead, with the two outer bins holding $3.1$ to $5.5\,\%$ of the cases
at day~1 and $4.6$ to $5.2\,\%$ at day~10 against $4\,\%$ for a calibrated
ensemble; the day-1 tilt of Z500 towards the highest rank ($3.6$ against
$1.7\,\%$) is the signature of a low mean error rather than of a spread
deficit. At the stations (Supplementary Fig.~\ref{fig:supp-rank-hist}) both
variables are mildly U-shaped, the residual under-dispersion of
Figure~\ref{fig:stations}c: the outer bins hold $6$ to $8\,\%$ for T2M, and
for WS10 the observation falls below every member in $6.6\,\%$ of the cases
at day~1.


\suppnote{Regional station verification}{note:regional}

\begin{figure}[!htb]
	\centering
	\includegraphics[width=\linewidth]{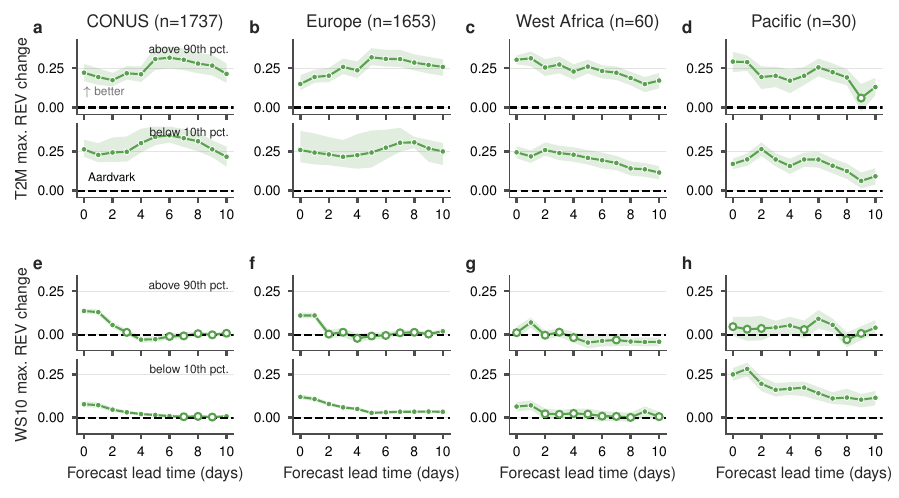}
	\caption{Regional station decision value: the statistic of
		Figure~\ref{fig:skill}b, the difference in maximum relative economic
		value (REV) between the E2E ensemble of Figure~\ref{fig:stations} and
		Aardvark, at HadISD stations per region of the Aardvark evaluation
		protocol (columns; verified station counts in the titles) for T2M
		(\textbf{a}--\textbf{d}) and WS10 (\textbf{e}--\textbf{h}), for
		events above each station's 90th (upper sub-panels) and below its 10th
		(lower sub-panels) climatological percentile; the dashed
		zero line marks Aardvark.}
	\label{fig:supp-regional}
\end{figure}

Supplementary Figure~\ref{fig:supp-regional} resolves the station comparison of
the main text per region, over the four regions of the upstream Aardvark
evaluation protocol (CONUS, Europe, West Africa, Pacific), for the end-to-end
finetuned ensemble of Figure~\ref{fig:stations} (Supplementary
Note~\ref{note:joint}). The decision value of the ensemble exceeds that of the deterministic
forecast significantly at 132 of the 176 (variable, region, tail, lead) points
and falls significantly below it at 7 (windy WS10 events at mid-range leads in
West Africa and CONUS), including the two sparse sets (West Africa, $\sim$60
verified stations; Pacific, $\sim$30), where the confidence intervals widen as
expected; for T2M the
margin is largest at the sparse networks, where the deterministic forecast has
little value beyond day~1 (West Africa, warm tail: peak REV $0.37$ vs.\
$0.06$ at day~1, $0.29$ vs.\ $0.03$ at day~5) while the ensemble retains a
third of the perfect-forecast value through day~5. For WS10 the ensemble's
advantage is confined to the first days in every region ($+0.07$ to $+0.14$ in
the windy tail and $+0.07$ to $+0.28$ in the calm tail at analysis time and
day~1), beyond which neither system retains much value for station wind
extremes: in CONUS, Europe and West Africa the peak REV of both is at most
$0.08$ from day~5. The Pacific set is the exception, where the ensemble keeps
calm-tail value through day~10 ($0.12$ against $0.01$) and windy-tail value at
day~5 ($0.17$ against $0.15$). The
decision-value gain is thus a property of the forecast system rather than of
any single, densely observed network, even where the ensemble mean is not the
more accurate forecast.

The station ensemble-mean RMSE comparison is shown in Supplementary Fig.~\ref{fig:supp-regional-rmse}.
In the two dense networks the ensemble mean is as accurate as Aardvark or better: over CONUS it is
significantly more accurate at every lead for both variables ($2$ to $12\,\%$
lower RMSE for T2M, $13$ to $18\,\%$ for WS10), and over Europe from day~3 for
T2M and day~6 for WS10 ($7\,\%$ lower at day~10 for both), with WS10 worse by up
to $6\,\%$ at days~1 to 3. Over West Africa the T2M RMSE of the two systems is
indistinguishable at most leads, while WS10 is $8$ to $16\,\%$ lower at every
lead. The Pacific set ($\sim$30 stations) is the exception: there the ensemble
mean is significantly worse for T2M at days~2 to 4 (up to $13\,\%$) and for WS10
at most leads to day~6 (up to $11\,\%$).

\begin{figure}[!htb]
	\centering
	\includegraphics[width=\linewidth]{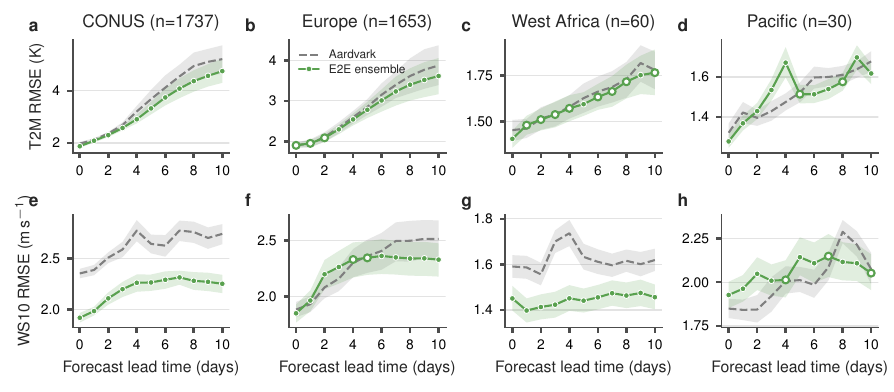}
	\caption{Regional station accuracy: ensemble-mean RMSE at HadISD
		stations per region of the Aardvark evaluation protocol (columns;
		median verified station counts in the titles) for T2M
		(\textbf{a}--\textbf{d}) and WS10 (\textbf{e}--\textbf{h}), for the
		E2E ensemble of Figure~\ref{fig:stations} (green) and Aardvark (grey
		dashed).}
	\label{fig:supp-regional-rmse}
\end{figure}

\suppnote{Relative economic value}{note:value}

\begin{figure}[!htb]
	\centering
	\includegraphics[width=\linewidth]{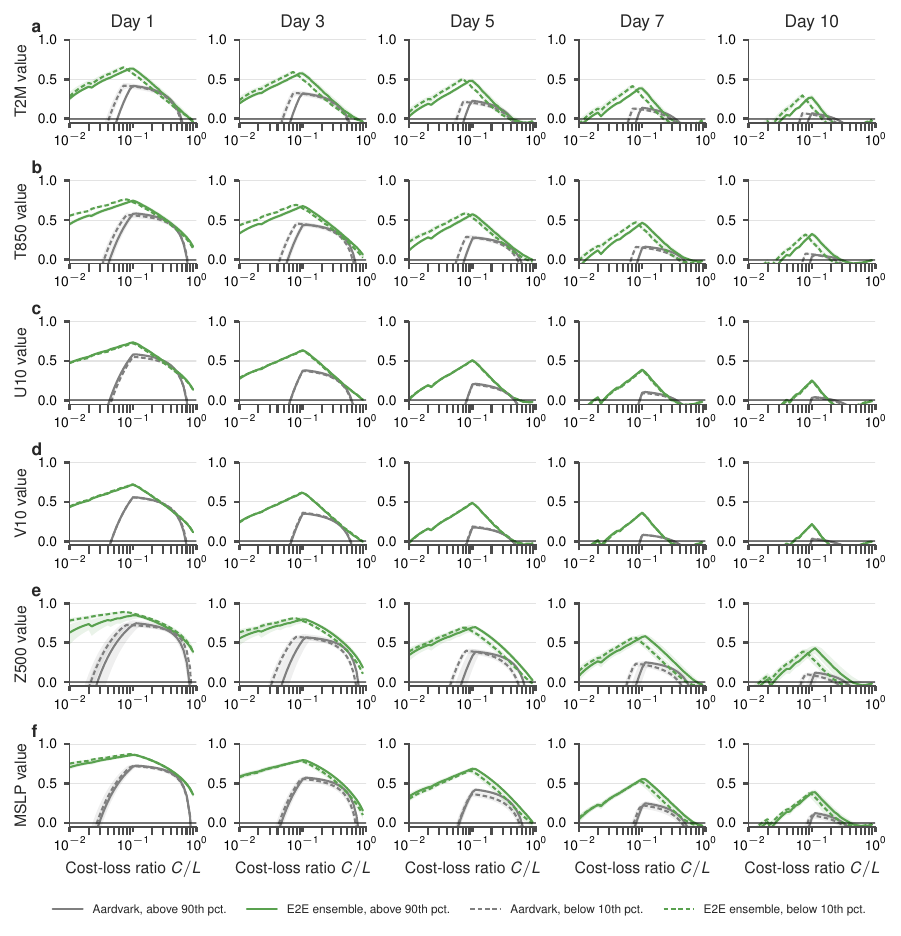}
	\caption{Relative economic value\cite{richardson2000value} on the grid
		as a function of the cost--loss ratio $C/L$ for events above the local
		90th (solid) and below the 10th (dashed) climatological
		percentile of each headline variable (rows \textbf{a}--\textbf{f}) at
		forecast days 1, 3, 5, 7 and 10 (columns). Green: the E2E ensemble as
		issued; grey: Aardvark, whose single yes/no call gives one curve for
		all users. The value curves peak at $C/L$ equal to the climatological
		event frequency, the point summarized as the maximum-REV difference in
		Figure~\ref{fig:skill}b and Supplementary
		Figure~\ref{fig:supp-regional}.}
	\label{fig:supp-value-curves}
\end{figure}

Supplementary Figure~\ref{fig:supp-value-curves} gives the full relative
economic value curves behind the decision-value panel of Figure~\ref{fig:skill}b
and the regional summary of Supplementary Figure~\ref{fig:supp-regional}. At day~1 the deterministic
forecast has positive value only for cost--loss ratios between about $0.06$ and
$0.56$, the ensemble for ratios between $0.01$ and $0.89$. Both curves peak at
the climatological event frequency, the point whose difference is summarized in
Figure~\ref{fig:skill}b.

\begin{figure}[!htb]
	\centering
	\includegraphics[width=\linewidth]{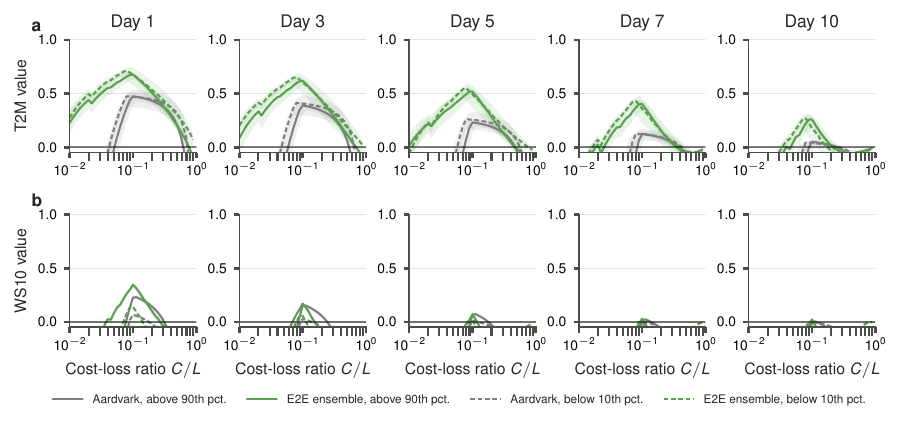}
	\caption{Relative economic value at HadISD stations as a function of
		the cost--loss ratio $C/L$ for T2M (\textbf{a}) and WS10 (\textbf{b})
		at forecast days 1, 3, 5, 7 and 10, over all verified stations, for
		events above each station's 90th (solid) and below its 10th (dashed)
		climatological percentile. Green: the E2E ensemble of
		Figure~\ref{fig:stations} as issued; grey: Aardvark.}
	\label{fig:supp-value-curves-station}
\end{figure}

At the stations, Supplementary Figure~\ref{fig:supp-value-curves-station} shows
the same curves for the end-to-end finetuned ensemble of
Figure~\ref{fig:stations}. For T2M the ensemble has the larger peak value at
every lead time in both tails, $0.68$ to $0.71$ at day~1 and $0.26$ to $0.27$ at
day~10, against $0.47$ and $0.04$ to $0.05$ for the deterministic forecast. At
day~1 it has value for cost--loss ratios from $0.01$ to $0.71$ (warm tail) and
$0.89$ (cold tail), against $0.06$ to $0.56$ and $0.05$ to $0.71$; from day~7
its value shifts to lower cost--loss ratios, and the deterministic forecast
keeps a small value above about $0.3$ in the warm tail. For WS10 the value of
both systems is small and confined to the first days. In the windy tail the
ensemble leads at day~1 (peak $0.34$ against $0.23$, with value for cost--loss
ratios from $0.04$ to $0.25$ against $0.08$ to $0.28$) and both systems are
without value from day~7; in the calm tail the ensemble leads at day~1 ($0.14$
against $0.07$) and both are without value from day~5.

\suppnote{Joint end-to-end finetuning against gridded and station targets}{note:joint}
\linelabel{rev:R2-5-supp}%

\paragraph*{Protocol.}
Both objectives start from the separately finetuned chain and train all encoder
(including the noise network) and processor weights on one rollout from lead~0
to day~10, with seven members per step, each with a fresh encoder draw
$\mathbf{a}$ and fresh dropout masks $\boldsymbol{\omega}$ ($p=0.05$). The loss
is the fair CRPS per lead, weighted by the inverse of its pre-finetuning value
so that short leads are not sacrificed to long ones; lead~0 anchors the
analysis. Training uses AdamW at a peak rate of $3\times10^{-6}$ with four
warm-up epochs and cosine annealing over 20 epochs. The \emph{gridded objective} scores the latitude-weighted
CRPS of the full gridded state against ERA5, with the encoder analysis as
lead~0, and trains no station decoder. The \emph{station objective} decodes each
lead with its own HadISD decoder, trained jointly, and scores the station CRPS
of one variable (one finetune each for T2M and WS10).

\paragraph*{Gridded objective.}
The gridded-objective finetune improves on the separate components at every lead
from day~1, as shown in Supplementary Fig.~\ref{fig:supp-joint-grid}: the variable-mean
fair CRPS of the six headline variables falls by $2.5\,\%$ at day~1, $11\,\%$ at
day~5 and $8\,\%$ at day~10, significantly at all 36 variable--lead
combinations, and the ensemble-mean RMSE is unchanged through day~1 and
$5$--$15\,\%$ lower from day~3. The day-1 over-dispersion of the separate chain
disappears (spread--skill ratio $1.03$ against $1.22$), the ratio stays within
$0.93$--$0.97$ from day~3, and the rank histograms are flatter at days~5 and 10
for every variable; the
analysis ensemble, in turn, becomes sharper (spread--skill ratio $0.60$ against
$0.76$ at lead~0, CRPS $+5\,\%$). The dropout fraction and encoder share remain
similar between the joint and separate components.

\paragraph*{Station objective.}
As shown in Supplementary Fig.~\ref{fig:supp-joint}, the station-objective
finetune lowers the T2M CRPS by $2$--$10\,\%$, significantly at all eleven
leads, and raises the spread--skill ratio at days~1 to 10 from $0.69$--$0.81$ to
$0.84$--$0.93$; the WS10 finetune lowers the CRPS by $19$--$29\,\%$. The larger
WS10 gain reflects the published WS10 station decoder, which shows strong under-dispersion
with the separate components (spread--skill ratio $0.32$--$0.41$ at days~1 to 10, against
$0.69$--$0.81$ for T2M), and the joint finetune, which retrains that decoder,
brings the ratio to $0.74$--$0.90$ and lowers the ensemble-mean RMSE by
$11$--$17\,\%$. The encoder share
rises by a few points at days~1 to 3 and the joint and separate components agree from day~5.

The joint objectives improve on the separately finetuned components, and both reproduce the dominance of
the dropout branch from day~1\linelabel{rev3:E-d}. Each objective improves only its own target. We therefore report each 
system on the target it was finetuned for.

\begin{figure}[!htb]
	\centering
	\includegraphics[width=\linewidth]{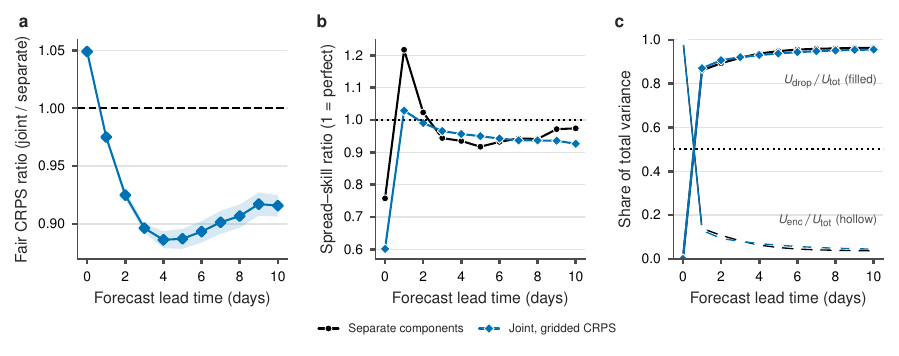}
	\caption{Joint finetuning versus separate finetuning on the grid
	(norm-conditioning chain), mean over the six headline variables. Black
	circles: separately finetuned components; blue diamonds: joint finetune
	against the gridded objective. \textbf{a}, Fair CRPS ratio to the
	separate components (below the dashed line favors the joint system),
	mean of the per-variable ratios. \textbf{b}, Spread--skill ratio.
	\textbf{c}, Dropout and encoder shares of the total variance,
	$U_{\mathrm{drop}}/U_{\mathrm{tot}}=\rho_{\mathrm{drop}}$ (filled
	markers) and $U_{\mathrm{enc}}/U_{\mathrm{tot}}$ (hollow markers),
	from the unbiased split of Eq.~\eqref{eq:anova}; lead~0 is the
	analysis, where the encoder share is 1.}
	\label{fig:supp-joint-grid}
\end{figure}

\begin{figure}[!htb]
	\centering
	\includegraphics[width=\linewidth]{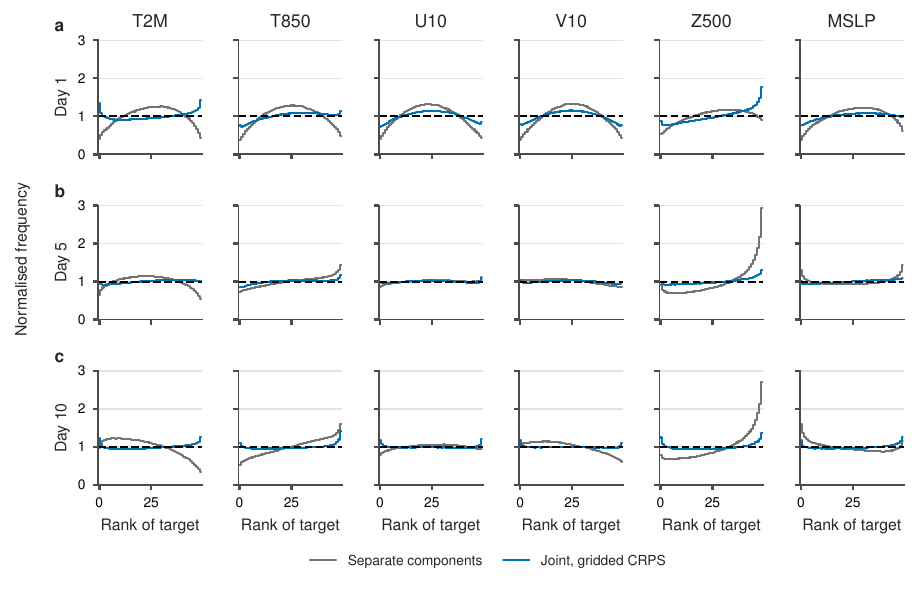}
	\caption{Gridded verification-rank histograms of the two systems of
		Supplementary Fig.~\ref{fig:supp-joint-grid}, per headline variable
		(columns) at days~1, 5 and 10 (rows). Grey: separately finetuned
		components; blue: joint finetune against the gridded objective; the
		dashed line marks a calibrated ensemble.}
	\label{fig:supp-rank-joint-grid}
\end{figure}

\begin{figure}[!htb]
	\centering
	\includegraphics[width=\linewidth]{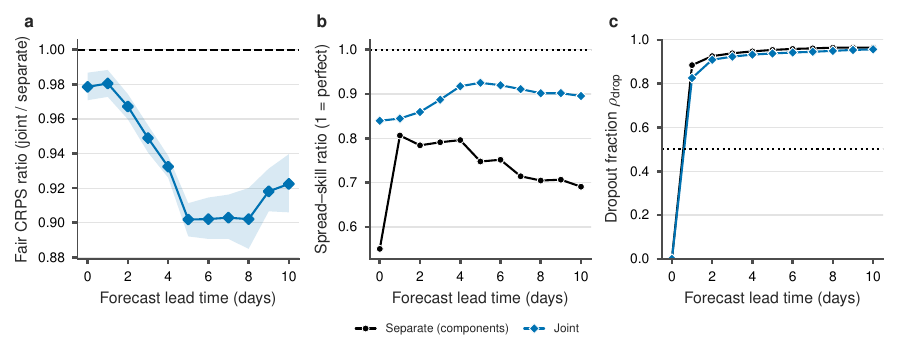}
	\caption{Station-objective joint finetuning versus separate finetuning
		(norm-conditioning chain of the main text), HadISD T2M. Black
		circles: separately finetuned components; blue diamonds: joint
		finetune with both sources active. \textbf{a}, Fair CRPS ratio to the
		separate components (below the dashed line favors the joint system).
		\textbf{b}, Spread--skill ratio. \textbf{c}, Dropout variance fraction
		$\rho_{\mathrm{drop}}$ (Eq.~\eqref{eq:anova}); lead~0 is the
		analysis, where $\rho_{\mathrm{drop}}=0$.}
	\label{fig:supp-joint}
\end{figure}

\suppnote{Notation}{note:notation}

Supplementary Table~\ref{tab:supp-notation} collects all symbols used throughout
the paper, grouped by the role they play: the forecast system and its
components, data and states, the two random sources, the nested ensemble, the
variance decomposition, and the verification statistics.

\begin{table}[!htb]
	\centering
	\caption{Symbols used throughout the paper.}
	\label{tab:supp-notation}
	\begin{tabular}{ll}
		\toprule
		Symbol                                                                                          & Meaning                                                                                                 \\
		\midrule
		\multicolumn{2}{l}{\emph{Forecast system}}                                                                                                                                                                \\
		$f_\theta = D_\theta \circ P_\theta^{(\tau)} \circ E_\theta$                                    & deterministic Aardvark system\cite{allen2025aardvark}                                                   \\
		$E_\theta$, $P_\theta^{(\tau)}$, $D_\theta$                                                     & encoder; processor over $\tau$ daily steps; station decoder                                             \\
		$g(\mathbf{x};\,\mathbf{a},\boldsymbol{\omega})$                                                & stochastic forecast map (Eq.~\eqref{eq:functional})                                                     \\
		$\theta$                                                                                        & learned weights, fixed at inference                                                                     \\
		$\tau$                                                                                          & forecast lead time in daily steps                                                                       \\
		\midrule
		\multicolumn{2}{l}{\emph{Data and states}}                                                                                                                                                                \\
		$\mathbf{x}$                                                                                    & heterogeneous observations of one assimilation window                                                   \\
		$\mathbf{x}_s$                                                                                  & a single observation stream (withheld in the denial experiment)                                         \\
		$\mathbf{z}_0$                                                                                  & assimilated initial state                                                                               \\
		$\hat{\mathbf{y}}_\tau$;\ $y$                                                                   & forecast fields at lead $\tau$; a single scalar component thereof                                       \\
		\midrule
		\multicolumn{2}{l}{\emph{Random sources}}                                                                                                                                                                 \\
		$\mathbf{a}\sim\mathcal{N}(\mathbf{0},\mathbf{I})$                                              & encoder-noise draw (observation-side source)                                                            \\
		$\boldsymbol{\mu}_\theta(\mathbf{x})$, $\boldsymbol{\sigma}_\theta(\mathbf{x})$                 & learned mean and noise amplitude of the stochastic encoder                                              \\
		$\boldsymbol{\omega}\sim q(\boldsymbol{\omega})$                                                & MC-dropout masks of one rollout (model-side source)                                                     \\
		$q(\boldsymbol{\omega})$                                                                        & dropout-mask distribution\cite{gal2016dropout}                                                          \\
		$p$                                                                                             & dropout rate ($p=0.05$)                                                                                 \\
		$\mathbf{e}$                                                                                    & per-patch noise embedding, MLP image of $\mathbf{a}$ (Supplementary Fig.~\ref{fig:supp-variants}a)      \\
		\midrule
		\multicolumn{2}{l}{\emph{Nested ensemble}}                                                                                                                                                                \\
		$M$,\ $N$                                                                                       & outer encoder draws / inner dropout rollouts per draw ($M{=}N{=}7$)                                     \\
		$K = MN$                                                                                        & total member count ($K=49$)                                                                             \\
		$\hat{y}^{(i,j)}$                                                                               & member $j$ of encoder draw $i$                                                                          \\
		$\bar{y}^{(i)}$;\ $\bar{y}$                                                                     & mean over the $N$ members of draw $i$; grand mean                                                       \\
		\midrule
		\multicolumn{2}{l}{\emph{Uncertainty decomposition}}                                                                                                                                                      \\
		$U_{\mathrm{tot}}$, $U_{\mathrm{enc}}$, $U_{\mathrm{drop}}$                                     & predictive variance and its encoder / dropout components (Eq.~\eqref{eq:ltv})                           \\
		$U_{\mathrm{enc}}\!\approx\!U_{\mathrm{alea}}$, $U_{\mathrm{drop}}\!\approx\!U_{\mathrm{epis}}$ & approximate aleatoric / epistemic reading (\hyperref[sec:vardecomp]{Methods})                           \\
		$\mathrm{MS}_{\mathrm{W}}$,\ $\mathrm{MS}_{\mathrm{B}}$                                         & within- and between-group mean squares (Eq.~\eqref{eq:ms})                                              \\
		$\widehat{U}_{\mathrm{enc}}$, $\widehat{U}_{\mathrm{drop}}$                                     & unbiased ANOVA estimates of the branch variances (Eq.~\eqref{eq:anova})                                 \\
		$\rho_{\mathrm{drop}}$                                                                          & dropout fraction $\widehat{U}_{\mathrm{drop}}/(\widehat{U}_{\mathrm{enc}}+\widehat{U}_{\mathrm{drop}})$ \\
		\midrule
		\multicolumn{2}{l}{\emph{Verification and statistical inference}}                                                                                                                                         \\
		$n$                                                                                             & number of forecast initializations in the test year ($n=337$)                                           \\
		$r_1$                                                                                           & lag-1 autocorrelation of a per-initialization score series                                              \\
		$\ell$                                                                                          & data-adaptive correlation length: bootstrap block length and maximum lag of the significance test       \\
		\bottomrule
	\end{tabular}
\end{table}

\end{document}